\documentclass[a4paper,11pt]{article}
\pdfoutput=1 
\usepackage{jheppub} 
\usepackage[T1]{fontenc} 
\usepackage[title]{appendix}
\usepackage[italian,english]{babel}
\usepackage{hyperref}
\hypersetup{
	colorlinks=true,
	linkcolor=blue,
	filecolor=magenta,      
	urlcolor=blue,
}
\usepackage{ifpdf}
\usepackage{subfigure}
\usepackage{tikz}
\usepackage[compat=1.1.0]{tikz-feynman}
\usetikzlibrary{arrows,shapes}
\usetikzlibrary{trees}
\usetikzlibrary{matrix,arrows} 				
\usetikzlibrary{positioning}				
\usetikzlibrary{calc,through}				
\usetikzlibrary{decorations.pathreplacing}  
\usepackage{pgffor}							

\usetikzlibrary{decorations.pathmorphing}	
\usetikzlibrary{decorations.markings}
\tikzset{
	vector/.style={decorate, decoration={snake}, draw},
	fermion/.style={draw=black, postaction={decorate}}, 
	scalar/.style={dashed,draw=black, postaction={decorate}}}
\tikzstyle{block} = [draw, rectangle, 
minimum height=3em, minimum width=6em]

\usepackage{amssymb}
\usepackage{amsfonts}
\usepackage{epsf}
\usepackage{rotating}
\usepackage{graphicx}
\usepackage{amsmath}
\usepackage{fancyhdr}
\usepackage{lineno}
\usepackage{babel}
\usepackage{graphics}
\usepackage{pstricks}
\usepackage{color}
\usepackage{multirow}
\usepackage{array}
\usepackage{diagbox}
\usepackage{cancel}

\usepackage[normalem]{ulem}
\usepackage{color}
\definecolor{darkgreen}{cmyk}{1,0,1,0.4}

\newcolumntype{P}[1]{>{\centering\arraybackslash}p{#1}}
\newcolumntype{M}[1]{>{\centering\arraybackslash}m{#1}}

\newcommand{\lsim}{\mathrel{\mathop{\kern 0pt \rlap
			{\raise.2ex\hbox{$<$}}}
		\lower.9ex\hbox{\kern-.190em $\sim$}}}
\newcommand{\gsim}{\mathrel{\mathop{\kern 0pt \rlap
			{\raise.2ex\hbox{$>$}}}
		\lower.9ex\hbox{\kern-.190em $\sim$}}}

\newcommand{\be}{\begin{equation}}
\newcommand{\ee}{\end{equation}}
\newcommand{\bea}{\begin{eqnarray}}
\newcommand{\eea}{\end{eqnarray}}
\def\barr{\begin{array}}
\def\earr{\end{array}}

\newcommand{\dis}{\displaystyle}

\newcommand{\gev}{\; {\rm GeV} }
\newcommand{\tev}{\; {\rm TeV} }

\def\gev{\ensuremath{\mathrm{\,Ge\kern -0.1em V\,}}}
\def\tev{\ensuremath{\mathrm{\,Te\kern -0.1em V\,}}}

\title{\boldmath Semi-Annihilation of Fermionic Dark Matter}
\author[a]{Priyotosh Bandyopadhyay,}
\author[b]{Debajyoti Choudhury,}
\author[c]{Divya Sachdeva}

\affiliation[a]{Indian Institute of Technology Hyderabad, Kandi,  Sangareddy-502287, Telengana, India}
\affiliation[b]{Department of Physics of Astrophysics, University of Delhi, Delhi 110007, India.}
\affiliation[c]{Laboratoire de Physique Th\'eorique et Hautes \'Energies (LPTHE),
UMR 7589 CNRS and Sorbonne Universit\'e, 4 Place Jussieu, F-75252 Paris, France}
\emailAdd{bpriyo@phy.iith.ac.in} 
\emailAdd{debchou.physics@gmail.com } 
\emailAdd{divyasachdeva951@gmail.com }

\preprint{ IITH-PH-0003/22}

\date{}
\begin{document}

\abstract{ The continued non-observation of events emanating from dark
  matter (DM) annihilations in various direct and indirect
  detection experiments calls into question the mechanism for
  determining the relic density of a weakly interacting massive
  particle.  However, if the relic density is determined
    primarily by a semi-annihilation process, as opposed to the
    usual annihilation, this tension can be ameliorated. 
 Here, we investigate a $Z_3$ symmetric effective
  field theory incorporating a fermionic dark matter that
  semi-annihilates to right-handed neutrinos (RHN).  
  The dynamics of the RHN and the impact of its late
  decays are also scrutinised while obtaining the correct DM relic.
  Finally, indirect detection bounds on the semi-annihilation
  cross-sections are drawn from the gamma-ray observations in the
  direction of Dwarf Spheroidal Galaxies (Fermi-LAT), and 
  including the projections obtained for the H.E.S.S. and the CTA
  detectors.}

	
\maketitle
\flushbottom
\section{Introduction}
\label{sec:introd}
Despite its stupendous success, the Standard Model (SM) leaves many
questions unanswered, and, in this article, we aim to address two of
them in an unified way. One of these pertains to the existence of Dark
Matter (DM) and the other to the problem of neutrino masses. While
both the questions have been addressed independently, and, on
occasions, even together, we consider a mechanism for establishing the
DM relic density that is relatively less
explored. Simultaneously, this mechanism would be seen to
naturally yield light neutrino masses of the right orders,
accommodating equally well both the normal and inverted hierarchy.

A popular mechanism for generating neutrino masses is to invoke the
Type-I seesaw mechanism via the introduction of the right-handed
neutrinos (RHNs) \cite{Minkowski:1977sc, Mohapatra:1979ia,
Mohapatra:1980yp, Schechter:1980gr, Yanagida:1980xy}. Being SM gauge
singlets, these rarely play any discernible role in laboratory physics
other than to generate neutrino masses and mixings and, hence, this
sector can hardly be probed, whether directly or indirectly. Attempts
have been made to ascribe to these a role in determining the DM relic
density, whether through freeze-out or the freeze-in
mechanisms~\cite{Falkowski:2009yz,Gonzalez-Macias:2016vxy,Escudero:2016ksa,Tang:2016sib,
Campos:2017odj,Batell:2017rol,Blennow:2019fhy,Hall:2019rld}. In such freeze-out scenarios, the DM particles pair-annihilate primarily to
these RHNs which, thereafter, decay to SM particles. Since the RHN has
to be lighter than the DM for this to take place, unless the DM itself
is very heavy, correct neutrino phenomenology would demand that the
neutrino-sector Yukawa couplings be small, thereby delaying the decay
of the RHN and, hence, the freeze-out
epoch~\cite{LFZ1,LFZ2,LFZ3}. Analogous effects are evinced in the
freeze-in mechanism as well \cite{Bandyopadhyay:2020qpn}.

The situation takes an interesting turn if the symmetry ensuring the
absolute stability of the DM is not the simple and popular $Z_2$.  A
more complicated symmetry (such as the $Z_3$ or $Z_5$) may dictate
that the the leading interaction is not a pair-annihilation of the DM
(whether directly into SM particles or other exotics), but one where,
say, three DM particles ${\cal D}$ must be
involved~\cite{Belanger:2012zr,Choi:2016hid}. This could manifest
itself in two different ways, a $ 3 {\cal D} \to \sum_\alpha {\cal
S}_\alpha$ annihilation (where ${\cal S}_\alpha$ denote $Z_3$-neutral
particles, whether within the SM or exotic) or of the form ${\cal D} +
{\cal D} \to {\cal D}^* + \sum_\alpha {\cal S}_\alpha$. The first
mechanism, seemingly, is very efficient in reducing the DM
density. It, however, is associated with low cross sections, simply on
account of the initial-state flux factors, and is relevant only for a
very light DM
particle~\cite{Choi:2016hid,Choi:2017mkk,Smirnov:2020zwf}. The second
process, on the other hand, reduces the number of DM particles by only
one, instead of the customary two in the case of the usual
pair-annihilation processes. With the ensuing slow-down of the
freeze-out process, the couplings required, in either case, to reach
the observed relic density~\cite{Planck:2018vyg} would be markedly
different leading to differing consequences for the direct or indirect
detection experiments.

The semi-annihilation process, as the leading mechanism, has been
studied in the context of both
scalar~\cite{semianniSca1,semianniSca2,semianniSca3,semianniSca4,semianniSca5,semianniSca6}
and vector~\cite{ semianniSca4, semianniVec1, semianniVec2,
semianniVec3} DM particles. For fermionic DM, though, such a
semi-annihilation process requires the participation of at least one
more fermion and, in a UV-complete theory, at least one additional
boson \cite{semianniFermi1, semianniSca4}. Gauge
invariance is most easily ensured if the the extra fermion in question is a
singlet under the SM and this brings into focus the possibility that
it is nothing but (one of) the aforementioned RHNs.

To this end, we concentrate on a fermionic DM ${\cal D}$ whose relic
abundance is determined primarily by semi-annihilation 
(${\cal D} \,
{\cal D} \to {\cal D}^c \, N_i$) into RHNs in a $Z_3$-symmetric
framework.  Rather than focus on a specific construction, we adopt a
model-independent approach focusing on effective operators.  Almost
irrespective of the UV-completion (we offer one in an appendix), the
RHNs further decay or annihilate into SM particles. The Yukawa couplings
responsible for the latter are allowed a wide range of values; in particular,
small values can delay the freeze-out with interesting cosmological
ramifications. 

It has been argued that models with DM annihilation to RHN are
severely constrained by direct and indirect
detection~\cite{Campos:2017odj,LFZ1} experiments.  We investigate
these aspects in the context of our model and find that while direct
detection constraints are very weak, the measurement of energetic
photon fluxes in the Fermi-LAT \cite{fermilat} and
H.E.S.S. \cite{HESS} experiments do serve to constrain the parameter
space. The viability is comparable to (or even better than)
models wherein a bosonic DM undergoes
semi-annihilation~\cite{semianniSca1}.

The rest of paper is planned as follows. In section \ref{model}, we describe
the scenario and delineate the  effective Lagrangian at the lowest order.
Section \ref{cosmoc} examines semi-annihilation as well as the other
relevant processes and thereby obtains the DM relic abundance for various
regions of the parameter space. Bounds on semi-annihilation cross-sections
 from FermiLAT and H.E.S.S, as well as sensitivity projections for the CTA are obtained  in
section \ref{Indirect}. We present our conclusions in
section \ref{conc}.  While a prospective UV completion is described in
Appendix \autoref{appendix:A}, calculational details are presented in
Appendices
\autoref{appendix:B} and \autoref{appendix:C}
respectively.

\section{The Model}
\label{model}
We augment the SM with two kinds of fermions, the RHN~\footnote{In
principle, we could do with just two $N_i$ fields as these would be
enough to generate two light neutrino masses.}  $N_i \, (i = 1, 2, 3)$
and the Dirac field $\chi$ for the
DM. While all the new fermions are
SM-singlets, we ascribe a nontrivial transformation of the
$\chi$-field under a $Z_3$, namely
\begin{equation}
{\cal S}_\alpha \to {\cal S}_\alpha \ , \qquad
N_i \to N_i \ , \qquad
\chi \to e^{2\pi i / 3} \chi \ .
\end{equation}

The $N_i$ may have Majorana masses as well as lead to Dirac masses
through Type-I Yukawa couplings. The corresponding Lagrangian may be
parameterised as
\begin{equation}\label{eq:L_N}
\mathcal{L_N} = \sum_i \overline{N_i} i \cancel \partial N_i
              + \sum_{i,j} (m_N)_{ij} \overline{N^c_i} N_j
              + \left[\sum_{i, k} (y_N)_{ik}\,\overline{L_i}\tilde{H}N_j + H.c.\right] ,
\end{equation}
where $L_i$ and $H$ are, respectively, the SM lepton doublets and the Higgs
field. Without any loss of generality, the (symmetric) Majorana mass
matrix $m_N$ may be considered to be diagonal.

After electroweak symmetry breaking, the usual Type-I seesaw mechanism
generates a mass matrix for the light neutrinos of the form
\begin{equation}
m_\nu\,=\, \frac{v^2}{2} \, y_N^T m_N^{-1} y_N,
\end{equation}
where $v=\,246\,\gev$. The matrix $m_\nu$, on diagonalization, would
yield the light neutrino masses and mixings. With the $N_i$ being
heavy, their mixings with the light species are tiny indeed. Nonetheless,
for $m_N \gsim 100\gev$, the small mixing still allows for prompt
decays\footnote{With $m_N$ being diagonal, flavour-changing decays are
suppressed. Even more suppressed are the lepton-number violating
decays.} such as $N_i \to \ell_i + W^+$ and $N_i \to \nu_i + Z/h$
with the branching fractions scaling approximately as $2:1:1$.
Since we would be typically interested in the range $100 \gev \lsim m_N
\lsim 1500\gev$,
the Yukawa couplings required to explain the neutrino masses would lie
in the range $ 10^{-7} \lsim y_N \lsim
10^{-6}$. Precise choices for the individual
couplings can always reproduce the observed masses and mixings, for
both normal and inverted hierarchies. However, since this is not the
primary goal, we desist from such an exercise, limiting
ourselves, instead, to admitting only such couplings as would produce
neutrino masses of the right order.

Given the particles and their quantum numbers, the DM field admits
only a free Lagrangian, namely
\begin{equation}
{\cal L}_\chi = \bar\chi \, (i \cancel\partial - m_\chi) \, \chi \ ,
\end{equation}
and the introduction of interaction terms would require either the
breaking of the $Z_3$ symmetry (which would destroy the stability of
the DM) or the presence of additional fields. Remaining agnostic to
the nature of such additional fields, except that these are heavy SM
singlets, one may, instead, express the
consequences of their inclusion in terms of an effective
theory. Restricting ourselves to the lowest dimensions, we have
\begin{equation}
{\cal L}_6 = \frac{c_{ab}}{\Lambda^2} \, {\cal O}_{ab}
= \frac{c_{ab}}{\Lambda^2} \, \left(\bar N \Gamma_a \chi \right)
                    \left(\bar{\chi^c} \Gamma_{b} \chi\right) + H.c. \ ,
\label{dim-6}
\end{equation}
where $\Lambda$ is the cutoff scale and we have dropped the index on
the $N$-fields. The Dirac matrices $\Gamma_a$ are to be chosen such
that Lorentz invariance is maintained. With
$\left(\bar{\chi^c} \gamma_\mu \chi\right)$ and
$\left(\bar{\chi^c} \sigma_{\mu\nu} \chi\right)$ vanishing
identically, the only remaining operators are\footnote{${\cal O}_{PP}$ (${\cal O}_{PS}$)
is nonvanishing as well, but leads to results identical to those for
${\cal O}_{SS}$ ${\cal O}_{SP}$. }
\begin{equation}
\begin{array}{lclclcl}
\mathcal{O}_{SS}&=& \dis \bar{\chi^c}\chi\;\bar{\chi}N, & &
\mathcal{O}_{SP}&=& \dis \bar{\chi^c}\chi\; \bar{\chi}\gamma_5 N, \\[1ex]
 \mathcal{O}_{AA}&= & \dis \bar{\chi^c} \gamma_{\mu}\gamma_{5}\chi \;
                  \bar{\chi}\gamma^{\mu}\gamma^{5}N, & \qquad \quad &  
 \mathcal{O}_{VA}&=& \dis \bar{\chi^c} \gamma_{\mu}\chi \; \bar{\chi} \gamma^{\mu}\gamma_5N \ .
\end{array}
\label{eq:oper}
\end{equation}
It might be argued, and rightly, that we have omitted other possible
terms such as $(\bar N \Gamma_a N) (\bar \chi \Gamma_b \chi)$ or even
analogous ones with the $N$ replaced by the SM fermion fields. With no
symmetry precluding their presence, such terms cannot be wished away
in their entirety.  However, the corresponding Wilson coefficients
could be small depending on the UV completion of the model. A trivial
solution would be to postulate that the coupling of the mediator to the
$N\chi$ current is much smaller than that to the $\chi\chi$ current.
One such completion is discussed at length in Appendix~\ref{appendix:A}.
However, independent of the details of the actual UV completion, we
assume that such terms ({\em viz.} $(\bar N \Gamma_a N)
(\bar \chi \Gamma_b \chi)$ {\em etc.}) are indeed suppressed (compared
to those in \autoref{eq:oper}), for a strong violation of such a
suppression would bring us back to the normal regime of the complete
annihilation process (as opposed to semi-annihilation) as the driving
force behind the determination of the relic density. This would render
operative the direct detection bounds (which, we will see, are
manifestly relaxed for semi-annihilation). In other words, we
deliberately choose to work in a regime where semi-annihilation
dominates.

The Lagrangian of \autoref{dim-6} would result in processes such as
$\chi\chi\chi \leftrightarrow N$ or $\chi\chi \leftrightarrow
N \chi^c$.  Of these, the $3\chi \to N$ process is naturally
suppressed on account of the small incident flux, while the decay
$N \to 3 \chi$ can proceed only if $m_N > 3 m_\chi$. Concentrating on
the $2 \to 2$ scattering
\begin{equation}
\chi(q_1) + \chi(q_2) \to N(q_3) + \chi^c(q_4) 
\label{the_process}
\end{equation}
the amplitude, then, reads,
\[
{\cal M}
= \dis 
\frac{c_{ab}}{\Lambda^2} \,
\Big[ (\bar u_3 \, \Gamma_a \, u_4)\, (u^T_2 \, \widehat \Gamma_b \, u_1)
  + (\bar u_3 \, \Gamma_a \, u_2) \, (u^T_1  \, \widehat \Gamma_b \, u_4)
  + (\bar u_3 \, \Gamma_a \, u_1) \, (u^T_4 \, \widehat \Gamma_b \,u_2) \Big]
\]
where $\widehat \Gamma_b \equiv \dis C \, \Gamma_b -
(C \, \Gamma_b)^T$ with $C$ being the charge conjugation matrix. Once
again, we see explicitly that $\Gamma_b = \gamma_\mu, \sigma_{\mu\nu}$
would lead to vanishing matrix elements. Henceforth, we make the
further simplifying assumption that only one of the remaining $c_{ab}$ is
non-zero and, for convenience, scale it to unity.  One might worry
about the renormalization group evolution of the Wilson
coefficients. However, unless the hidden sector is a strongly
interacting one, this evolution is not expected to be any different
from that in an usual ultraviolet complete theory of DM. In
particular, within the standard freeze-out paradigm, it is only within
a small range of momentum scales that most of the interesting physics
happens, and the Wilson coefficients do not vary appreciably over such
a range.


\section{Semi-annihilation  and the DM relic abundance}
\label{cosmoc}
Having set up the effective theory, we must now validate it against
observations. To this end, we begin with the cosmological constraints
emanating from the PLANCK observations of temperature and polarization
anisotropy in the cosmic microwave background, mainly, the relic
abundance of DM, $\Omega_{DM}h^2 =
0.1199\pm0.0012$ \cite{Planck:2018vyg}.

As it will turn out, the mechanism of interest is the freeze-out of
the DM.  In this model, it proceeds via its semi-annihilation to RHNs
through $\chi \chi \rightarrow \chi^c N$. The Yukawa coupling of the
RHNs allows these to decay through a multitude of channels, viz
\[
N_i\to \ell_i\; W,\qquad N_i\to Z\; \nu_i \quad {\rm and} \quad N_i\to
     h \nu_i.
\]
Although the first two of the decay modes proceed only through the
mixing of the $N_i$ with the SM neutrinos, it is easy to see that the
leading modes in each subclass have similar partial widths.

\begin{figure}[h]
	\centering \includegraphics[width=0.70\textwidth]{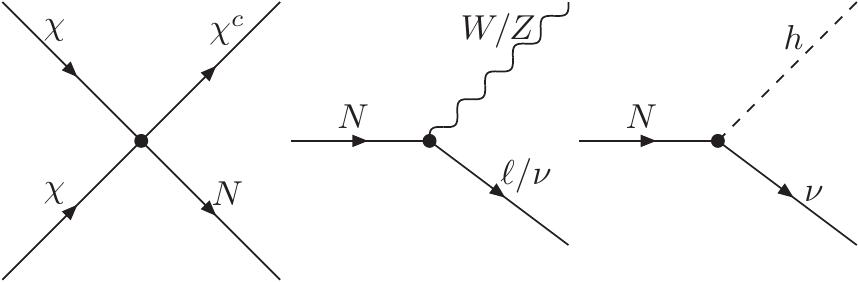} \caption{\em
        Feynman diagram of the semi-annihilation dark matter to RHN
        and the decays of RHN. } \label{Feyn}
\end{figure}

With \autoref{eq:L_N}\,\&\,\autoref{dim-6} encapsulating all of the
interactions of the $\chi$ and the $N_i$, their densities are
essentially governed by the processes in \autoref{Feyn}. The
consequent Boltzmann equations are best represented as the evolution
of the {\em yield} $Y$ ($Y \equiv n/s$ with $n$ being the number
density of the species under consideration and $s$ the ambient entropy
density) with the inverse of temperature, {\em viz.} $x \equiv m_\chi
/ T$. These can be summarised as (see Appendix \ref{appendix:C} for a
derivation)
\begin{equation}
\label{eq_Boltz1}
\barr{rcl}
\dis
\frac{dY_{\chi}}{dx}& =&
\dis
\frac{- x\,s(m_\chi)}{2H(m_\chi)}\sum_i\langle\sigma v\rangle^{\chi \chi \rightarrow \chi N_i} \left[
        Y_{\chi}^2- \frac{Y^{Eq}_{\chi}}{Y^{Eq}_{N}}
        Y_{\chi}\,Y_{N_i}\right], \\[3ex]
\dis \frac{dY_{N_i}}{dx}& =&
\dis \frac{x\,s(m_\chi)}{2H(m_\chi)} \langle\sigma v\rangle^{\chi \chi \rightarrow \chi N_i} \left[Y_{\chi}^2-\frac{Y^{Eq}_{\chi}}{Y^{Eq}_{N}} Y_{\chi}\,Y_{N_i}\right]
	-\frac{\Gamma_{N_i}\,x}{H(m_\chi)} (Y_{N_i}-Y_{eq}) \ .
\earr
\end{equation}
Here, the thermally averaged cross-section~\cite{DEramo:2017ecx} is
given by
\be
\barr{rcl}
\dis \langle\sigma(\text{ij}\rightarrow \text{kl})\,v\rangle
& = & \dis \left[4\,K_2(m_i/T)K_2(m_j/T) \, m_i^2 \,
m_j^2\,T\right]^{-1}
\\[2ex]
& \times & \dis \int^\infty_{E^{\text{min}}_{\rm cm}}dE_{\rm
 cm} \, \lambda(E^2_{\rm
 cm},m_i,m_j) \, \sigma_{\text{ij}\rightarrow \text{kl}}(E_{\rm
 cm}) \, K_1(E_{\rm cm}/T)
\earr     
\label{eq:vsigma}
\ee
where $E_{\rm cm}$ is the center of mass energy, $K_n$ is the modified
Bessel function of the second kind of order $n$ and $\lambda(a,b,c)$
is the usual K\"allen function.  The expressions for the cross
sections and the decay widths of the RHN can be gleaned from
Appendix \ref{appendix:B}.

\begin{figure}[h]
	\centering \mbox{\subfigure[]{\includegraphics[width=0.50\textwidth]{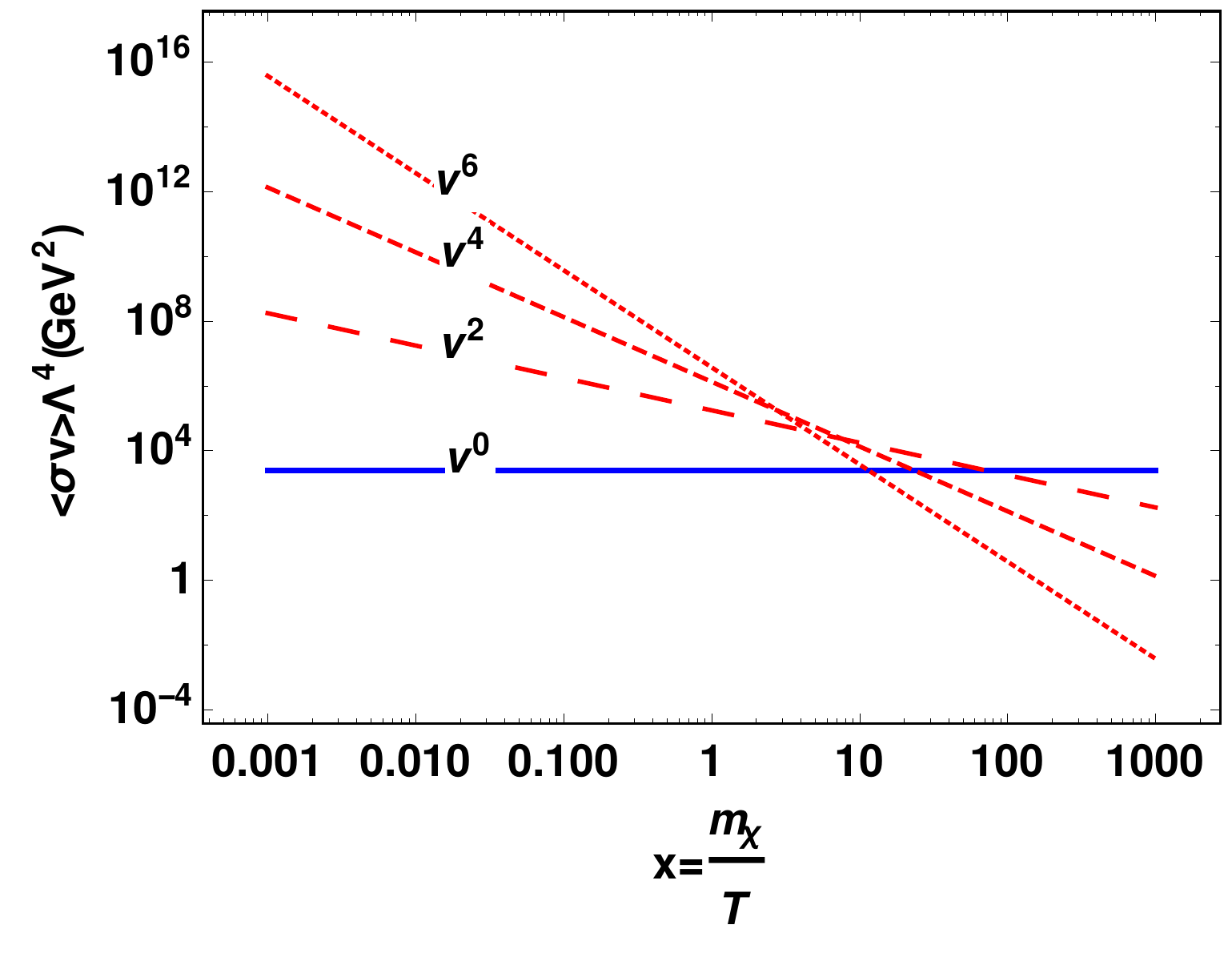}} \hfill \subfigure[]{\includegraphics[width=0.50\textwidth]{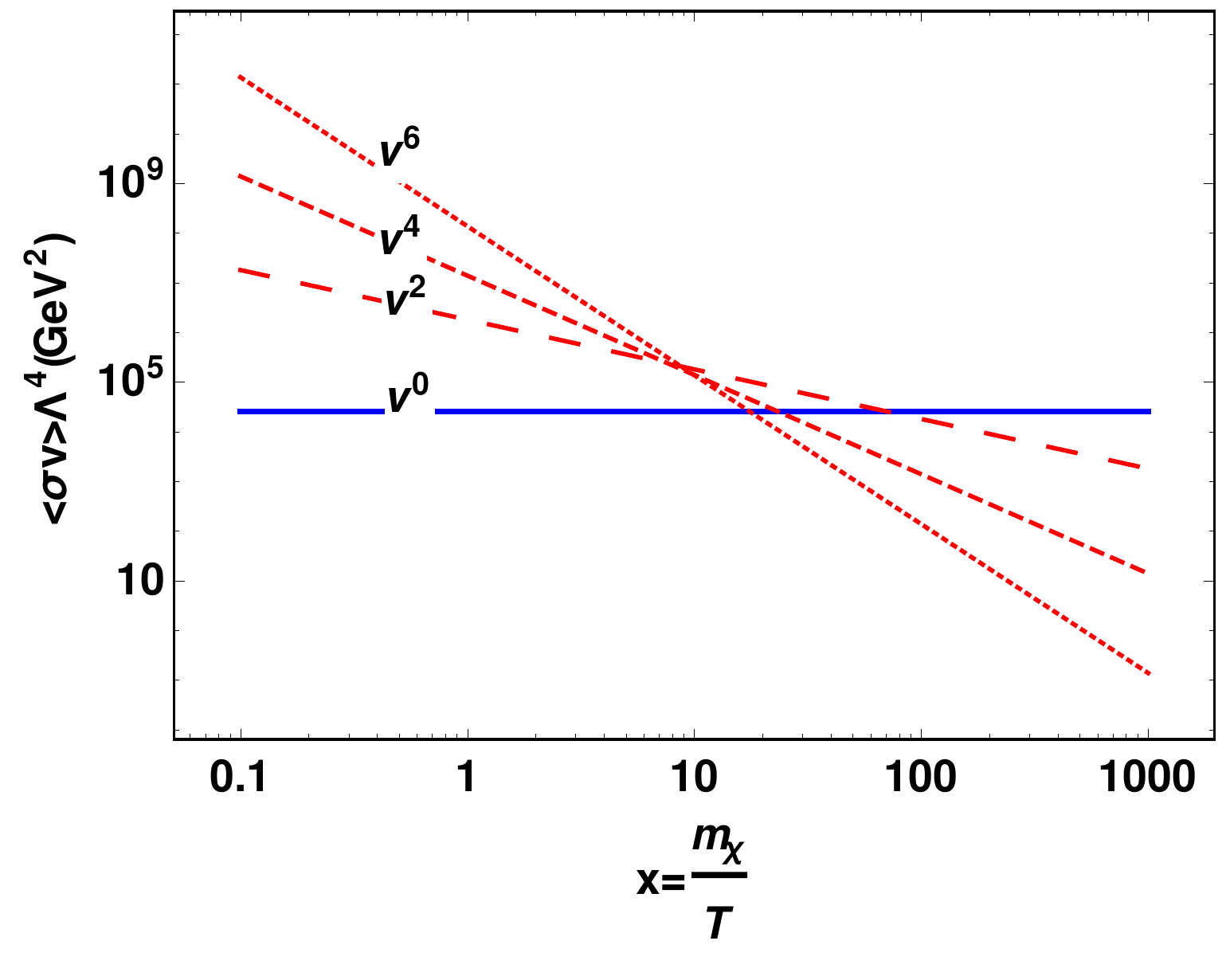}}} \caption{\em
        Different terms in expansion of $\langle\sigma
        v\rangle\times\Lambda^4$ plotted as a function of
        $x=m_{\chi}/T$ for SS interaction with a) $m_\chi = 300 {\rm
        GeV}, m_N = 100 {\rm GeV}$ b) $m_\chi = 1000 {\rm GeV}, m_N =
        100 {\rm GeV}$. } \label{fig:sigmaVexp}
\end{figure}
A particular aspect needs to be appreciated at this juncture. It is
commonplace to write $\langle \sigma v \rangle$ as a power series in
the relative velocity $v$. In most scenarios, the terms involving the
higher powers fall off quickly, allowing one to understand the
scattering in terms of $s$-wave or $p$-wave dominance etc. In the
present situation though, while the terms do continue to fall off
similarly (see \autoref{fig:sigmaVexp}), additional complications
arise on account of the peculiarity of the effective interaction (and,
thus, the amplitude). No single partial wave dominates over the entire
range of evolution; rather, different terms can be important during
the decoupling of DM, with the relative importance being a function of
both the DM and the RHN masses. For example,
as \autoref{fig:sigmaVexp} demonstrates, for a larger $m_\chi / m_N$,
the $v^2$ term falls faster with $m_\chi / T$ whereas the $v^6$ term
falls slower.  Consequently, we use the full integral
in \autoref{eq:vsigma}. The yield of dark matter, as of today, is well
approximated by $Y_\chi(\infty)$ and can be used to calculate the
relic abundance using
\[
\Omega_\chi h^2 = \displaystyle\frac{m_\chi
s_\infty Y_\chi(\infty) h^2}{\rho_c} \ ,
\]
where $s_{\infty} = 2889.2\, \rm{cm}^{-3}$ and
$\rho_c\,=\,1.05 \,\times\, 10^{-5} h^2\, \text{GeV}^2 \text{
cm}^{-3}$.
\begin{figure}[!tb]
	\centering \mbox{\subfigure[]{\includegraphics[scale=.5]{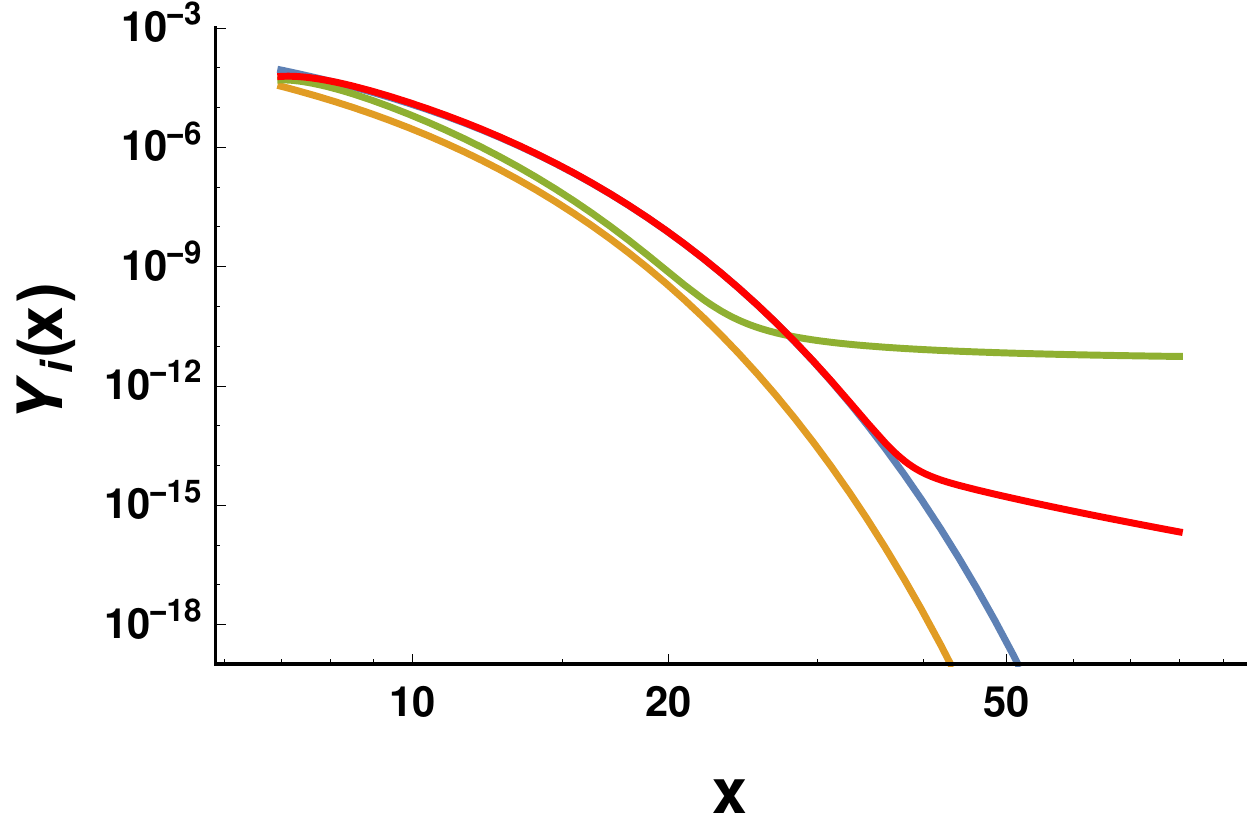}}\hspace{.25cm} \subfigure[]{\includegraphics[scale=.5]{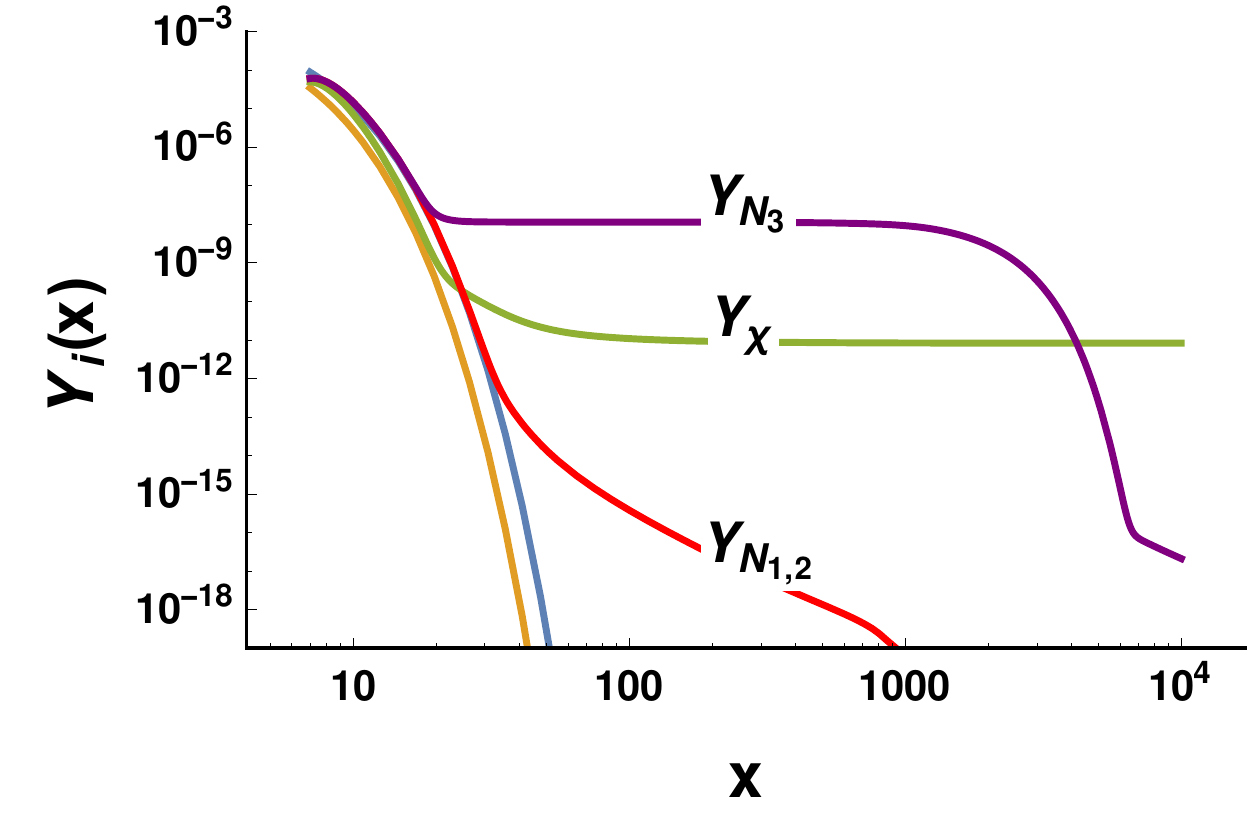}}} \mbox{\subfigure[]{\includegraphics[scale=.5]{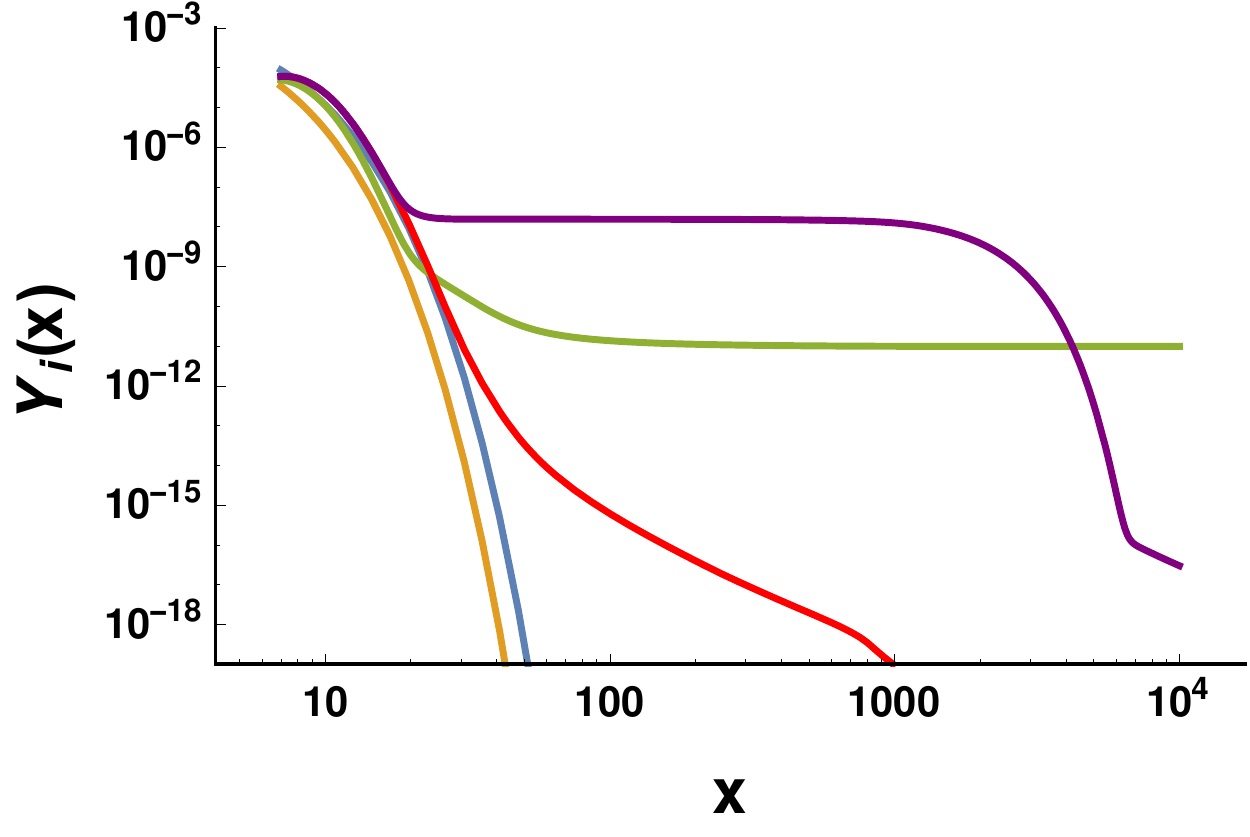}}\hspace{.25cm} \subfigure[]{\includegraphics[scale=.5]{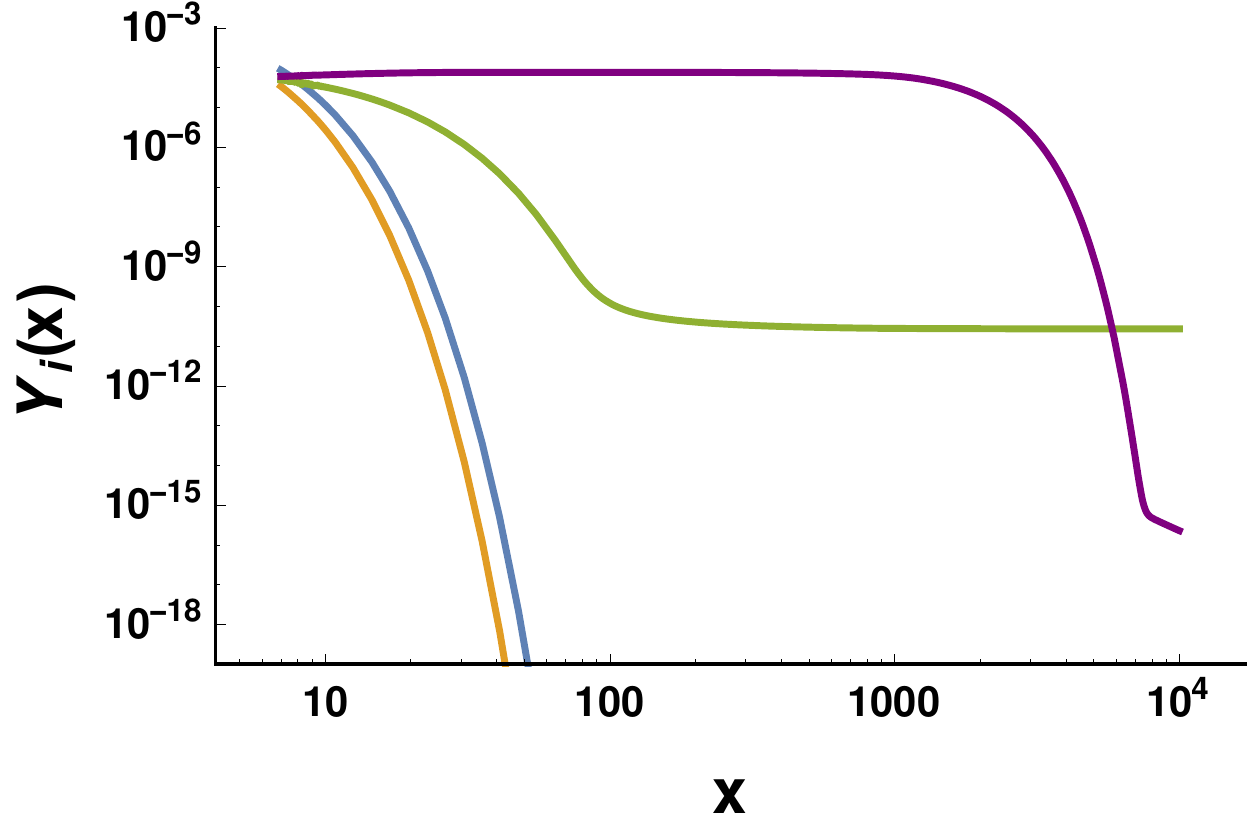}}} \caption{\em
	Yield of $\chi$ and right handed neutrino for $m_\chi =
	120, \Lambda = 500, m_N = 100$ as a function of
	$x\,=\,\frac{m_{\chi}}{T}$. The yellow and blue curves show
	the {\em extrapolated} equilibrium abundances of the DM and
	the RHN respectively. The green and red (purple) curves show
	the {\em actual} evolution of abundance of DM and RHN for
	different cases of Yukawa coupling: a) $y_{N_i} =
	(10^{-7},10^{-7},10^{-7})$, b) $y_{N_i} =
	(10^{-7},10^{-7},10^{-10})$, c) $y_{N_i} =
	(10^{-7},10^{-10},10^{-10})$ and d) $y_{N_i} =
	(10^{-10},10^{-10},10^{-10})$. The red and purple curves in b)
	and c) correspond to RHN with largest and smallest Yukawa
	couplings. } \label{fig:abundance}
\end{figure}

With the RHN playing a prime role in determining the DM relic
abundance, it is imperative to examine the evolution of their
density. The mass ratio $m_N/m_X$ as well as the size of the
couplings, both play important roles resulting in a somewhat
complicated behaviour as summarised in \autoref{fig:abundance}, Here,
the yields of the DM and the RHNs are shown as a function of
$x=\frac{m_{\chi}}{T}$ for $m_{\chi}=120, \, m_{N_i}=100$ GeV and
$ \Lambda = 500$ GeV. To highlight the individual epochs of freezing
out, we employ a ruse wherein the yellow and blue curves represent the
abundances of DM and RHN, respectively under the (unphysical)
assumption of these never having fallen out of equilibrium.  The
actual evolution for the DM are depicted by the green curves while
those for the RHN with $y_{N_i}=10^{-7}(10^{-10})$ are shown by the
red(purple) ones.

At temperatures of the order of the electroweak scale, all the SM
particles would, of course be in equilibrium.  For $m_N \sim {\cal
O}(100\, \mbox{GeV})$, a sizable $y_N \, (\sim 10^{-7})$ would ensure
that the aforementioned decays of the $N_i$, along with the reverse
processes, would keep these particles in equilibrium with the leptons
and the weak bosons, and, hence with the entire SM
sector. Furthermore, as the RHN are postulated to be lighter than the
$\chi$, a large enough $y_{N_i}=10^{-7}$ (red curve) for all three
RHNs, would imply that they would decouple at an epoch later than
$\chi$ (green curve) and, hence, would have a much smaller relic
density, as can be seen in \autoref{fig:abundance}(a).

The situation changes dramatically if the $y_{N_i}$ is/are much
smaller, as is illustrated by \autoref{fig:abundance}(b) with $y_{N_i}
= (10^{-7},10^{-7},10^{-10})$ and \autoref{fig:abundance}(c) with
$y_{N_i} = (10^{-7},10^{-10},10^{-10})$. While the densities of the
$N_i$ with a larger Yukawa coupling (red curve) continue to behave as
in
\autoref{fig:abundance}(a), those with small Yukawa coupling(s) (purple)
behave differently as their interactions with the SM sector are
now much weaker than the Hubble expansion rate. Hence, while the $N_i$
with larger Yukawas are still in equilibrium with the SM (and, through
$\chi \chi \to \chi N$, so is the DM), these other RHNs decouple
slightly before the DM does.  Consequently, post decoupling,
their number densities continue to be much larger than that of the
DM. This has an interesting consequence. Owing to the larger density
of such $N_i$, the process $\chi N_i \to \chi \chi$ would occur more
frequently than the reverse, and the DM density would be
slightly replenished before becoming constant at a
slightly lower temperature. A careful comparison of
\autoref{fig:abundance}(b) and \autoref{fig:abundance}(c) shows that
this effect is more pronounced in the latter, which is but a
consequence of the fact that the latter case corresponds to two $N_i$
with a small Yukawa coupling as opposed to just one for the former.

For the sake of illustration, in \autoref{fig:abundance}(d) we depict
the case where $y_{N_i} = (10^{-10},10^{-10},10^{-10})$, a situation
that cannot explain all the neutrino masses.  Owing to all the Yukawas
being tiny, the RHN (purple) and, hence, the DM (green) both decouple
from the equilibrium distribution very early. However, the tiny
Yukawas also mean that the RHN decay late. Consequently, the $\chi
N_i \to \chi \chi$ keeps on replenishing the DM relic abundance for a
longer period.

With the sizes of the Yukawa couplings having such marked imprints on
the thermal history, it is natural to expect that the relic densities
too would be quite different. However, the dependence is not too
severe as the value is primarily driven by $m_\chi$ and
$\Lambda$. Indeed, on the scale of
\autoref{fig:abundance}, the differences do not show up. A careful calculation,
though, shows that, in the four cases examined, $Y_\chi(x = 10^4)$
read $(a)\; 4.4\times 10^{-12}$, $(b)\; 8.15\times 10^{-12}$, $(c)\;
9.7\times 10^{-12}$ and $(d)\; 2.7\times 10^{-11}$, respectively.

We now turn to other consequences of the presence of the $N_i$.  First
and foremost, while their lifetimes are not ultrasmall, they still
decay fast enough for overclosure to be an issue. On the other hand,
their not so inconsiderable lifetimes could, presumably, alter physics
at the era of big bang nucleosynthesis (BBN) or even the CMBR
epoch. The latter alongwith the relative abundance of the light
elements (BBN) encode information about the thermal history of the
early universe, and is well described by SM physics. The decay of
particles post such epochs have the potential of destroying the
agreement between theoretical predictions and the
observations. Fortunately though, in the present context, the longest
lifetime of RHN is at most $0.001\,{\rm s}$ (corresponding to $y_N\sim
10^{-10}$ and $m_N\,=\,100\gev$) whereas BBN occured in the cosmic
time window $0.1{\rm s} - 10^{4}{\rm s}$ and CMB is even later.
Thus, the abundance of RHN would have depreciated to
a very large extent. To be quantitative, studies carried
out in ref.~\cite{Kawasaki:2017bqm} show that the BBN
constraints dictate that the fractional abundance
of particles of mass ${\cal O}(100 \gev)$ and decaying into
hadronically at $t \gsim 1 \, {\rm s}$ after the inflationary phase
should be less than $10^{-12}$. Thus, given the smallness of
the $N_i$ lifetimes, BBN constraints are trivially satisfied. Similar
is the case with the CMBR observations.

On the other hand, late decays of the $N$ would presumably leave their
mark in the sky. The energetic neutrinos can be coming from RHN decay
can be detected by experiments like SuperK, IceCube,
etc~\cite{Bandyopadhyay:2020qpn}. However, such bounds are found to be
much weaker compared to the ones coming from photon spectrum as
discussed in \autoref{Indirect}.

\subsection{Scale and DM mass dependencies:}
\begin{figure}[!tb]
\centering
\mbox{
\subfigure[]{\includegraphics[width=0.50\textwidth]{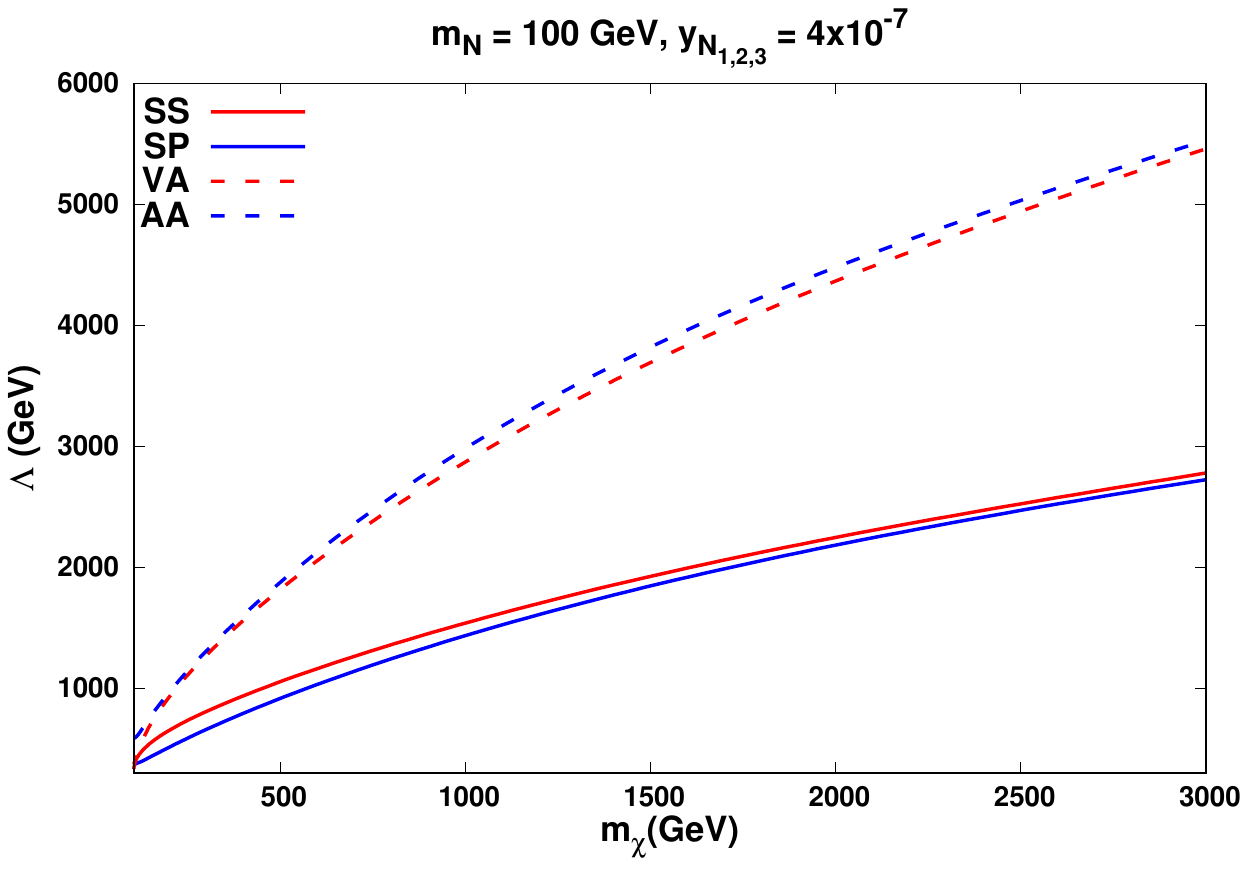}}
\subfigure[]{\includegraphics[width=0.50\textwidth]{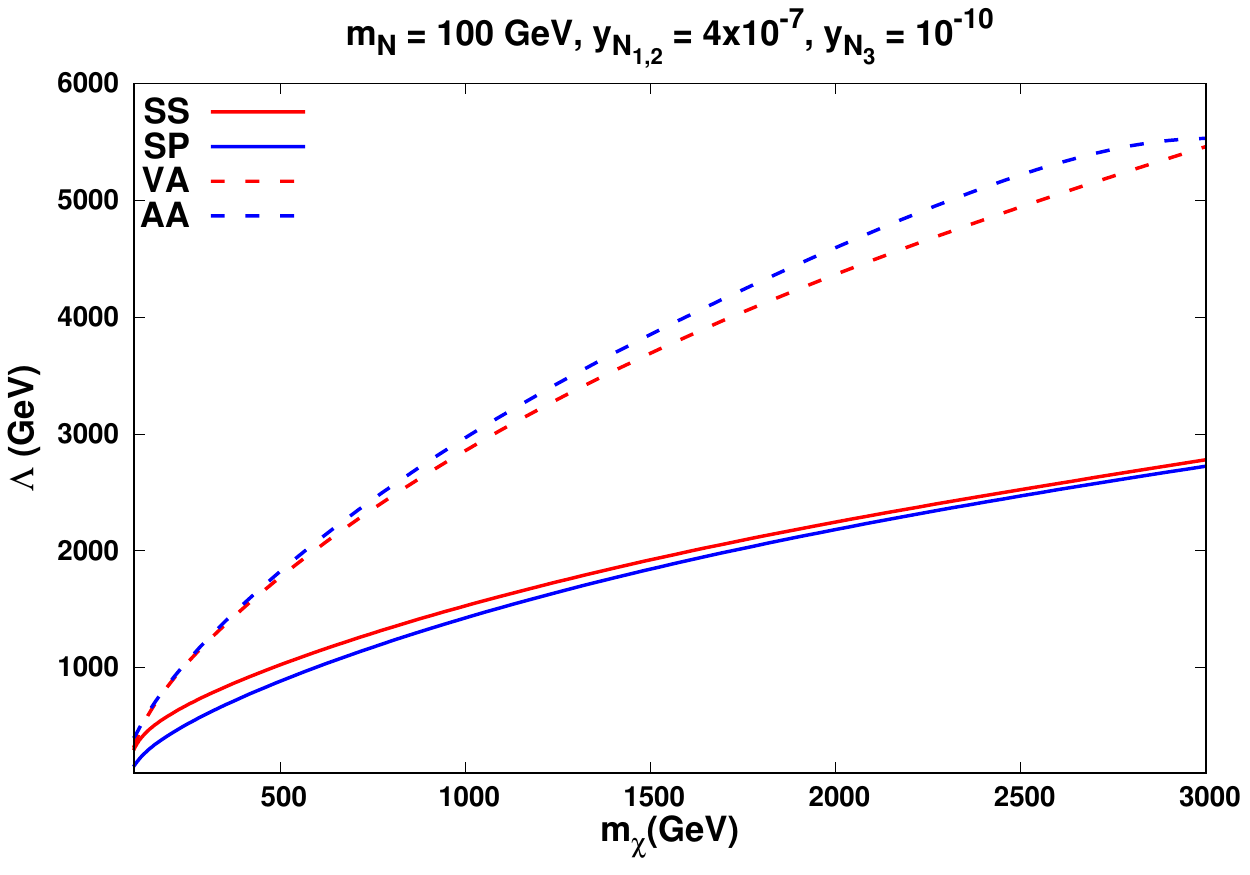}}
}
\mbox{
\subfigure[]{\includegraphics[width=0.50\textwidth]{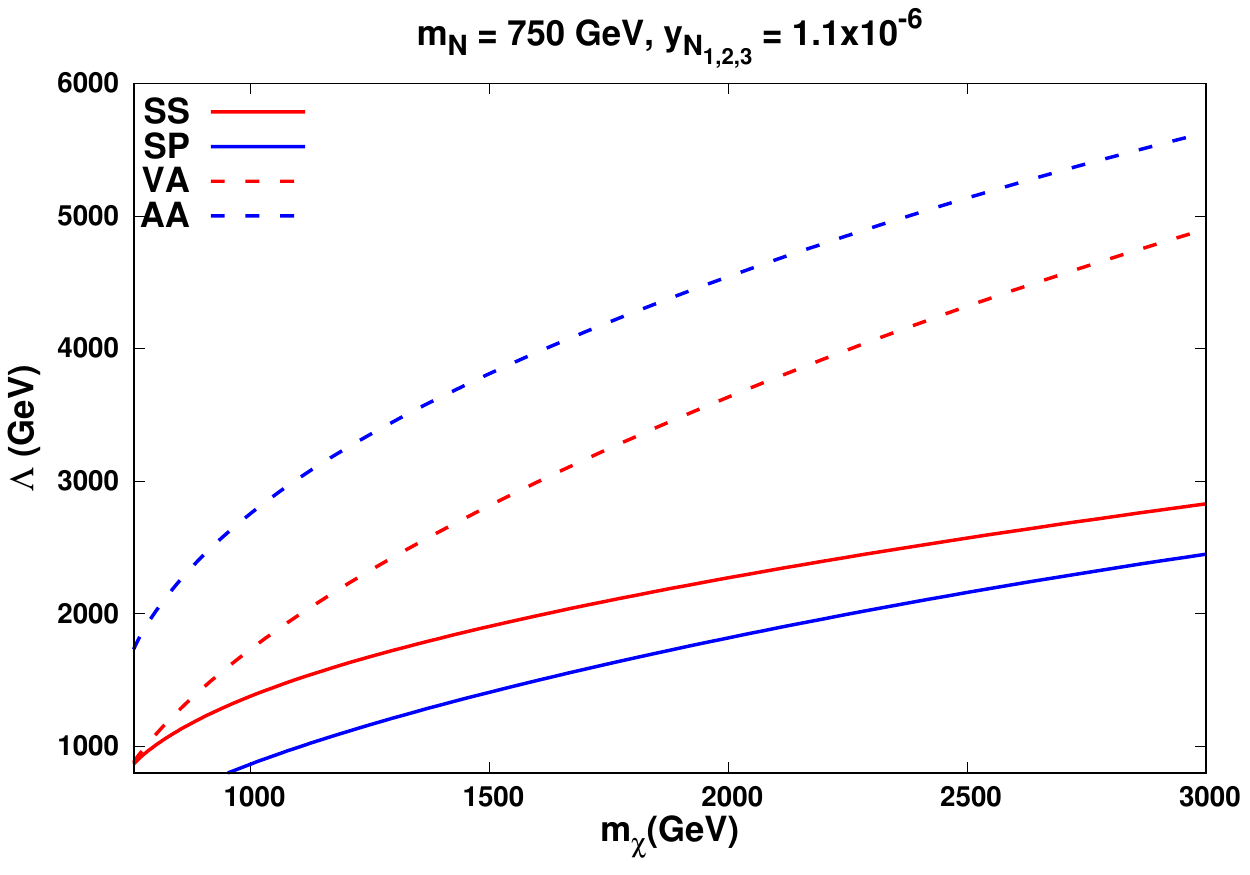}}
\hfill \subfigure[]{\includegraphics[width=0.50\textwidth]{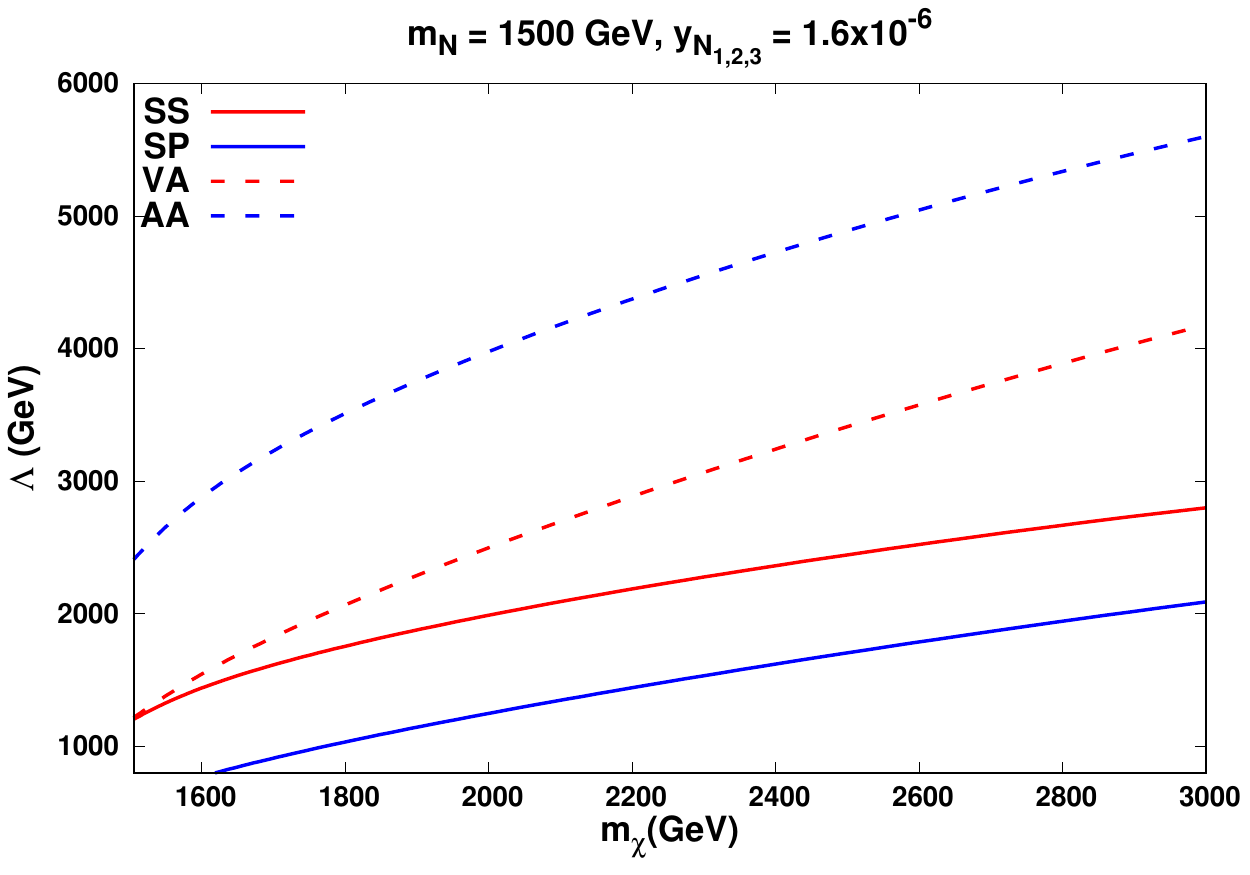}}
}
\caption{\em Contour on $m_\chi-\Lambda$ plane satisfying $\Omega
 h^2 = 0.1200\pm0.0012$ for different interactions for different
 values of $m_N$ and $y_N$. }
\label{fig:relic_results}
\end{figure}

Having studied the various
 dependencies, we now proceed to determining the parameter space
 corresponding to the observed relic density. The various panels
 of \autoref{fig:relic_results} depict the contours, in the
 $m_\chi-\Lambda$ plane satisfying $\Omega h^2 = 0.1199\pm0.0012$ as
 applicable for different Lorentz structures of the effective
 4-fermion interaction. Each panel corresponds to a different set of
 values for the Yukawa couplings $y_N$ with a common heavy
 neutrino mass $m_N$. Understandably, the area above the curves
 (larger $\Lambda$) corresponds to smaller interaction strengths and,
 hence, an overabundance of DM. On the other hand, the area below the
 curve leads to an under abundance and, hence, is still 
 allowed (though only at the cost of postulating an additional
 component of the dark matter).  Note that these results are obtained
 for $m_\chi\,>\,m_N$. For $m_\chi\,<\,m_N$, three body decays
 such as $N\to \nu +Z^*/h^*\to \nu \bar{f}f$ or $N\to \ell +
 W^* \to \ell \bar{f} f'$ would be important. However, we do not
 explore this in the present work.
 
The curves in \autoref{fig:relic_results} imply that obtaining the
correct amount of energy density requires the mass of dark matter
particles to increase with $\Lambda$. This could be easily understood
by looking at the functional dependencies of the cross sections.  With
the DM particle being non-relativistic at the epoch of decoupling, the
center of mass energy ($E_{\rm cm}$) for the process $\chi\chi\to\chi
N$ is close to $2m_\chi$ and the thermal average of the cross section
scales roughly as $m_\chi^2F(m_N/m_\chi)/\Lambda^4$ where
$F(m_N/m_\chi)\,<\,1$ is specific to the form of the interaction
Lagrangian. This immediately indicates that $\Lambda$ should scale
roughly as $m_\chi^2$ with deviations depending on the form of $F(m_N/m_\chi)$.
Furthermore, that the $SS \, (\equiv PP)$ and $SP$ structures require a cut-off scale $\Lambda$ to be nearly half
of what is required for $VA$ or $AA$ can be easily traced to the fact
that the two disparate sets of cross sections differ by a factor of nearly 16.

As for the $m_N$ dependence, note that for the VA and SP structures, $F(m_N/m_\chi)$ is a
monotonically
decreasing function of the argument. 
The consequent decrease in the total cross section with an increase in
$m_\chi$ would need to be compensated by an associated decrease in
$\Lambda$ so as to lead to an unchanged relic abundance, and this is
reflected in a comparison of panels $(a), (c)$ and $(d)$
of \autoref{fig:relic_results}.  On the other hand, for the SS
and AA cases, the cross sections do not change appreciably with
$m_N$. 
 
Finally, \autoref{fig:relic_results}(b) corresponds to the case where one of
 the RHN has small decay width ($y_{N_i}= 10^{-7}, 10^{-7},
 10^{-10}$), which leads to a nominally larger relic abundance (as shown
 in \autoref{fig:abundance}(b)) as compared to the case where
 all the RHNs have same Yukawa coupling, $y_{N_i}= 10^{-7}$. 

Before we end this section, it is contingent upon us to examine the validity of the effective field theory paradigm {\em vis \'a vis} the parameter space
required to satisfy the correct relic density. Clearly, the cut-off
scale $\Lambda$ for such a description must be larger than the masses
of any particle present in the theory. In the present case, this is
definitely satisfied by the entire parameter space for the VA and the
AA cases. For the (pseudo) scalar current-current structures, though,
this is not so for larger $m_\chi$ values (although, for smaller
$m_\chi$, this does hold). In other words, the effective theory
approach may not be entirely valid and the use of the full theory
seems to be called for.  It is, however, easy to ascertain that the
consequent changes in this part of the analysis would not be severe,
and that the results obtained herein are precise enough for an exploratory
study that this is.

\section{Constraints from indirect detection of DM}
\label{Indirect}
The semi-annihilations of the DM would also lead to electrons,
positrons, quarks, neutrinos, gamma-rays etc. with different energy
spectra. Of these, gamma-ray observations have been used prominently
to probe the presence of DM decay, annihilations and
semi-annihilations. The reasons for the choice are twofold:
$\gamma-$ray emissions from various astrophysical objects are well
understood, and unlike charged particles, photons travel through
interstellar and intergalactic matter suffering relatively little
obstruction. Thus, by comparing the expected gamma-ray flux to the
measured one, one can derive upper limits on DM interactions in the
absence of any excess.

The processes relevant to the present study is the semi-annihilation
of DM to RHN which, subsequently, decay via $W^\pm \ell,\,Z \nu$ and
$h\nu$ with the bosons cascading down to quarks (hadronizing almost
instantly) and leptons. The resultant charged particles from the decay
yield gamma rays, either during hadronization or as a result of final
state radiation processes. To this end, we use observations from the
following three experiment set-ups to constrain the effective
operators analysed in this work:
\begin{itemize}
  \item Fermi-LAT: The satellite covers a wide energy range, starting
    from $0.5$~GeV and going up to $0.5$~TeV.  To derive the exclusion
    limits, 6 years of data from 15 dwarf Spheroidals (see Table 1 of
    \cite{fermilat}), based on the \textbf{pass 8} event analysis
    \cite{fermilat} is used.
\item The High Energy Stereoscopic System (H.E.S.S.) is sensitive to
  high energy $\gamma-$rays spanning energies ranging from
  $0.2$ TeV to 30 TeV. The initial
    phase (H.E.S.S. I) had already published competitive limits on DM
    annihilations cross sections~\cite{HESS:2011zpk}. With the
    addition of a large dish at the center of the array, the second
    phase (H.E.S.S. II) is expected to lead to much stronger
    constraints~\cite{HESS}.  To exploit this expected sensitivity, we
    first tune our analysis to reproduce the H.E.S.S I
    data~\cite{HESS:2011zpk}, and follow it by rescaling it with the
    projected sensitivity in the $b \bar b$ channel as calculated
    in Ref.\cite{HESS}.
\item With more than a 100 individual telescopes, the Cherenkov
  Telescope Array (CTA) would
    be well-placed to detect high-energy gamma-rays over a very
    wide energy range ($\sim$60 GeV to $\sim$300 TeV)
  and is projected to be more than ten times as sensitive as the
    currently operating ones. This unprecedented sensitivity is
    expected to yield very strong constraints (if not an actual
    discovery) especially when looking at gamma-rays from Galactic
    center.
\end{itemize}
Note that the Fermi-LAT experiment is more sensitive to gamma-rays at
low energies, whereas H.E.S.S. and CTA are sensitive at larger
energies.

\subsection{The $\gamma$--ray spectrum:}
A gamma-ray signal from DM annihilation or semi annihilation is
inferred from the number of photons at a given energy bin from a
portion of the sky. The quantity that captures this information is the
differential flux, which is proportional to: $(i)$ the square of the
number density of the dark matter particle; $(ii)$ the
thermal-averaged product of the annihilation cross-section
and the velocity ($\langle\sigma v\rangle$);
$(iii)$ the number of photons produced per dark matter process as a
function of energy, i.e the energy spectrum (dN/dE); and $(iv)$ the
size and density of the region of the sky under study, as captured in
the $J$-factor. On inclusion of the appropriate normalization factor,
the differential flux for dark matter semi-annihilation is
\begin{equation}
	\frac{d\phi_\gamma}{dE}=\frac{1}{8\pi m_{\chi}^2} \langle \sigma v \rangle  \frac{dN_\gamma}{dE} J,
	\label{semiflux}
\end{equation} 
where
\begin{equation}
	J=\int_{l.o.s}\frac{ds}{r_\odot}\left(\frac{\rho(r(s,\theta))}{\rho_\odot}\right)^2,
\end{equation}
and includes an integration over the line-of-sight
between the observatory and the source. This factor obviously depends
on the dark matter distribution which we have assumed to follow the
NFW profile \cite{Navarro:2008kc}.

\begin{figure}[h]
	\centering
	\includegraphics[width=0.60\textwidth]{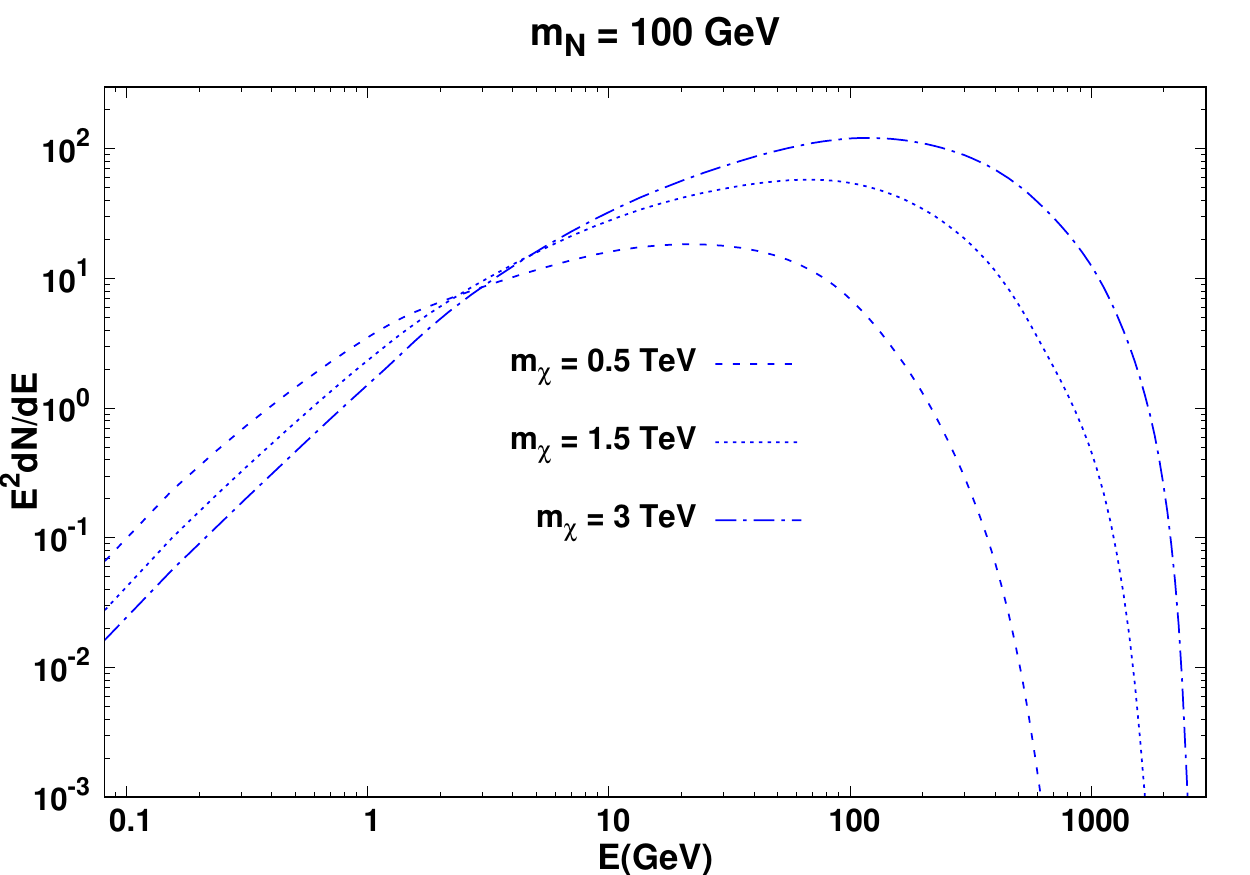}
	\caption{\em  Photon energy spectrum for different values of $m_\chi$ at a fixed value of $m_N\,=\,100$ GeV. \label{fig:spectrum_dm}}
\end{figure}

\begin{figure}[h]
\centering
\mbox{\subfigure[]{\includegraphics[width=0.50\textwidth]{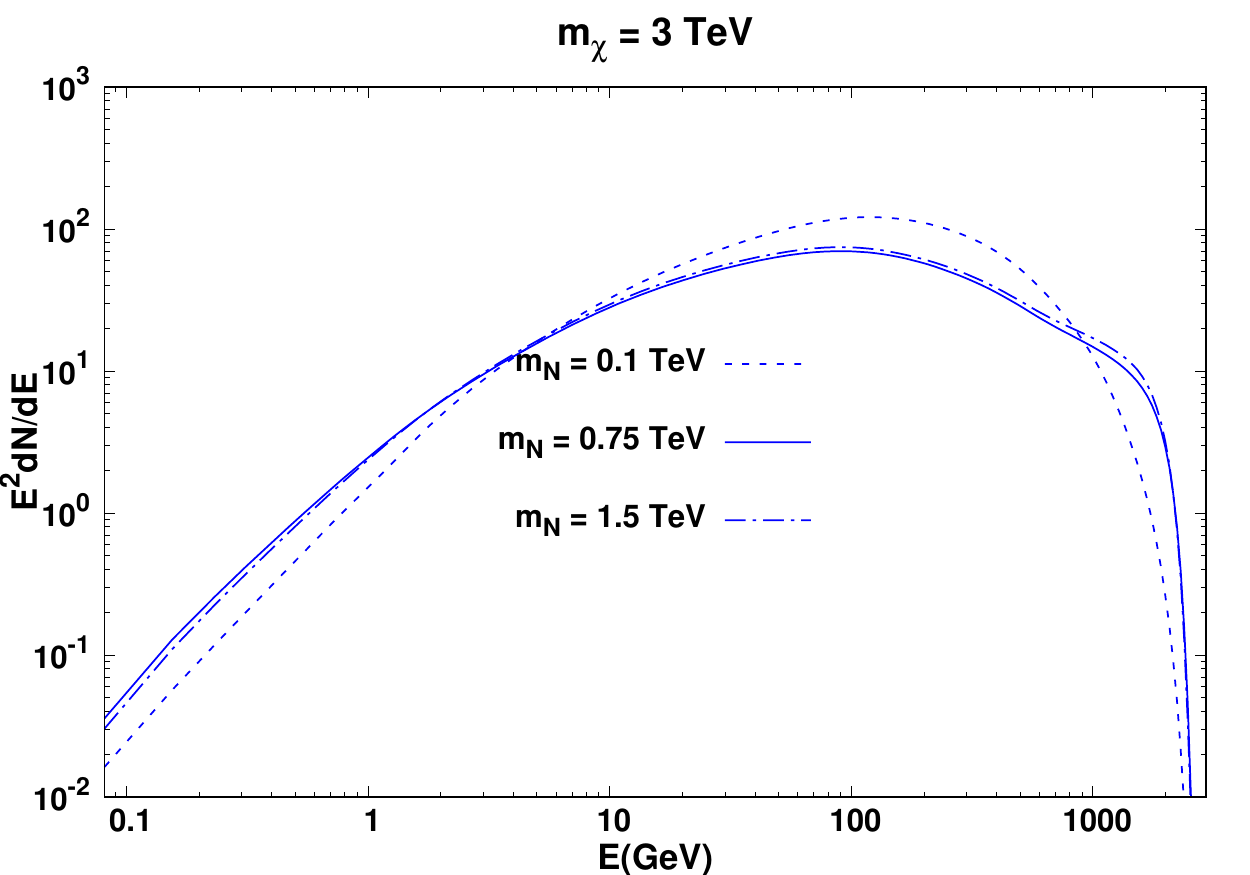}}
  \hfill
  \subfigure[]{\includegraphics[width=0.50\textwidth]{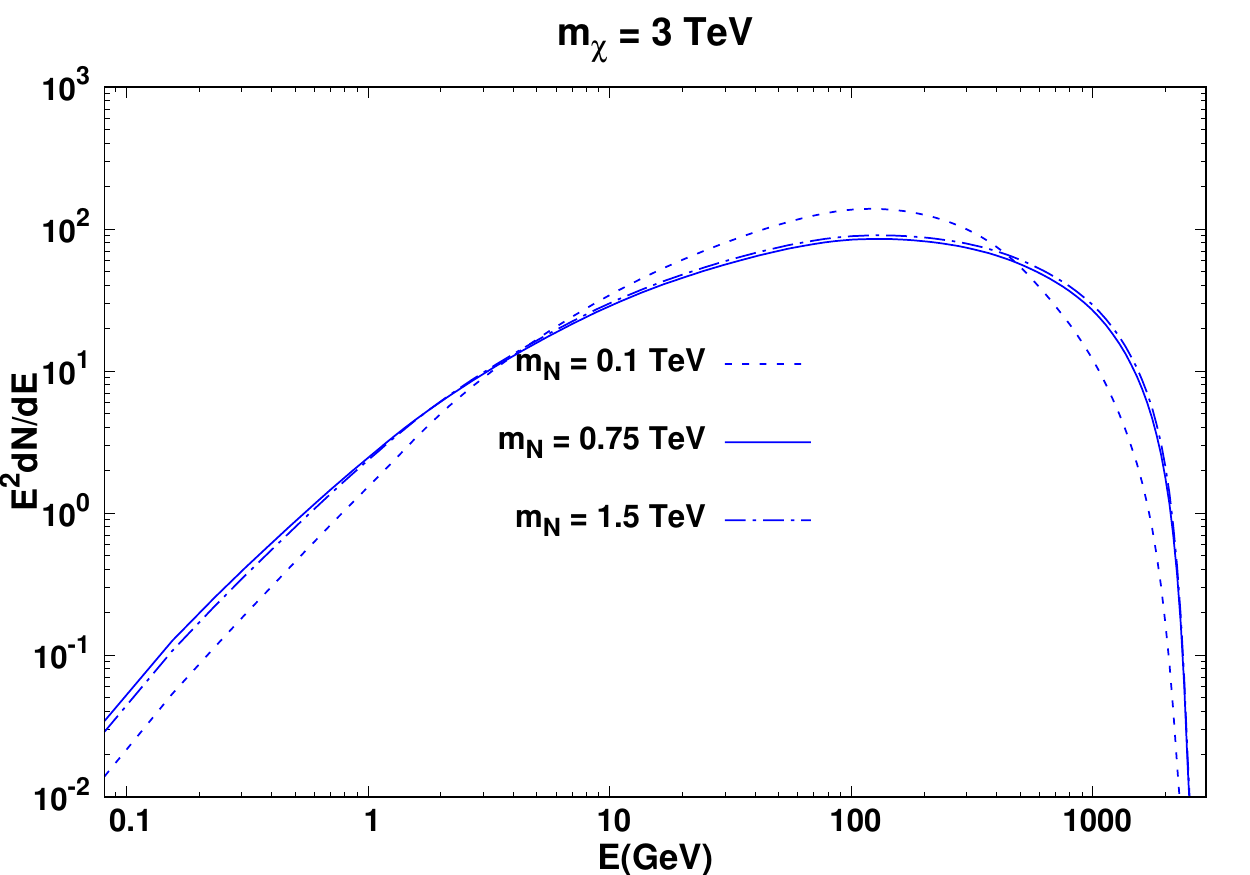}}}
\caption{\em Photon energy spectrum for $m_\chi\,=\,3$ TeV for
  different values of RHN mass and different final state leptons (a)
  electron and (b) tau-lepton.\label{fig:spectrum}}
\end{figure}

We compute the energy spectrum $dN/dE$
  due to the semi-annihilation processes\footnote{For the sake of simplicity,
    we do not include inverse Compton
  scattering. Inclusion of this would only increase the number of photons at
  larger energies and would improve limits obtained here. Our analysis
  here is thus a conservative one.}  {\em etc.},
  including the hadronization effects, using a suitably augmented
  version of PYTHIA8.3~\cite{Sjostrand:2014zea}.
    The resultant normalised energy
spectrum for different values of $m_\chi$ at a fixed RHN mass (with the RHN decaying to
  a final state containing an $e^\pm$) in
\autoref{fig:spectrum_dm}. Since the DM is non-relativistic, the
energy of the 
RHN can be estimated to be
\[E_N\,=\,\dis\frac{s+m_N^2-m_\chi^2}{2\sqrt{s}}\sim
\dis\frac{m_N^2+3m_\chi^2}{4m_\chi} \ .
\]
As expected, and as illustrated by \autoref{fig:spectrum_dm}, a
heavier DM leads to a more energetic RHN. With the extra energy being
transmitted to the $N$'s decay products, this would lead to a harder
gamma-ray spectrum. Given the expression for $E_N$, it is obvious that
it changes substantially with $m_N$ only when the latter is comparable
to $m_\chi$. This property is naturally transmitted to the $N$'s
daughters, including to the gamma-ray spectrum
(\autoref{fig:spectrum}). For a given $m_\chi$, the heaviest of the
$N$s would lead to the hardest $\gamma$--spectrum. On the other hand,
by virtue of being most boosted, the lighter ones lead to a larger
number of energetic photons (\autoref{fig:spectrum}). Finally, for a given $(m_\chi, m_N)$ combination, with each
being heavier than a 100 GeV, the leptons from the $N$'s decay would
have a virtually flavour-independent energy spectrum. Nonetheless, the
$\tau$-channel would result in more energetic photons owing simply to
its decay and subsequent hadronization. Understandably, the spectrum
for the muonic decay channel would lie in between the two.

\subsection{Results}
Armed with the discussions of the preceding subsection, we are now in
a position to compare the expectations of our model with the
observational data from Fermi-LAT and H.E.S.S. on the one hand, and
compare with the expected sensitivity of the CTA on the other. To this
end, we use the GamLike package \cite{GAMBIT} to compute the
likelihood function $\mathcal{L}(\hat{\mu},\hat{\theta}|\mathcal{D})$
where $\mathcal{D}$ represents the observational data, $\hat{\theta}$
the (unknown) parameters describing the background and $\hat{\mu}$ a
set of parameters encapsulating the features of the dark matter
model. The limits on the semi-annihilation cross-section are then
derived through the statistical test (TS),
\begin{equation}
  TS = -2 \ln{\left( \frac{\mathcal{L}(\hat{\mu}_0,\hat{\theta}|\mathcal{D})}
    {\mathcal{L}(\hat{\mu},\hat{\theta}|\mathcal{D})}
    \right)},
\end{equation}
with $TS>2.71$ corresponding to $95\%$~C.L. exclusion. 
Here, $\hat{\mu}_0$ corresponds to the null hypothesis,
{\em i.e.}, the absence of dark matter.

\begin{figure}[h]
\centering
\mbox{\subfigure[]{\includegraphics[width=0.50\textwidth]{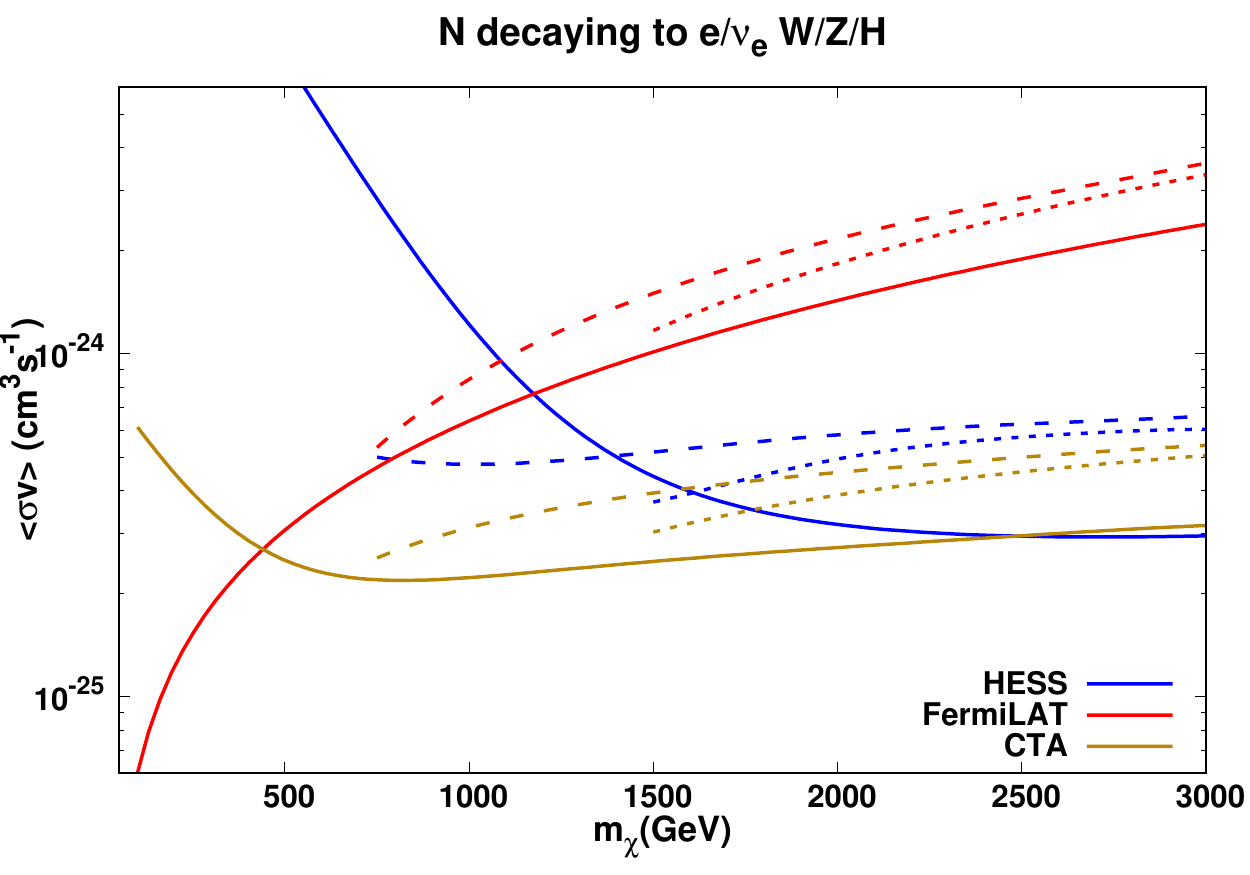}}
\hfill
\subfigure[]{\includegraphics[width=0.50\textwidth]{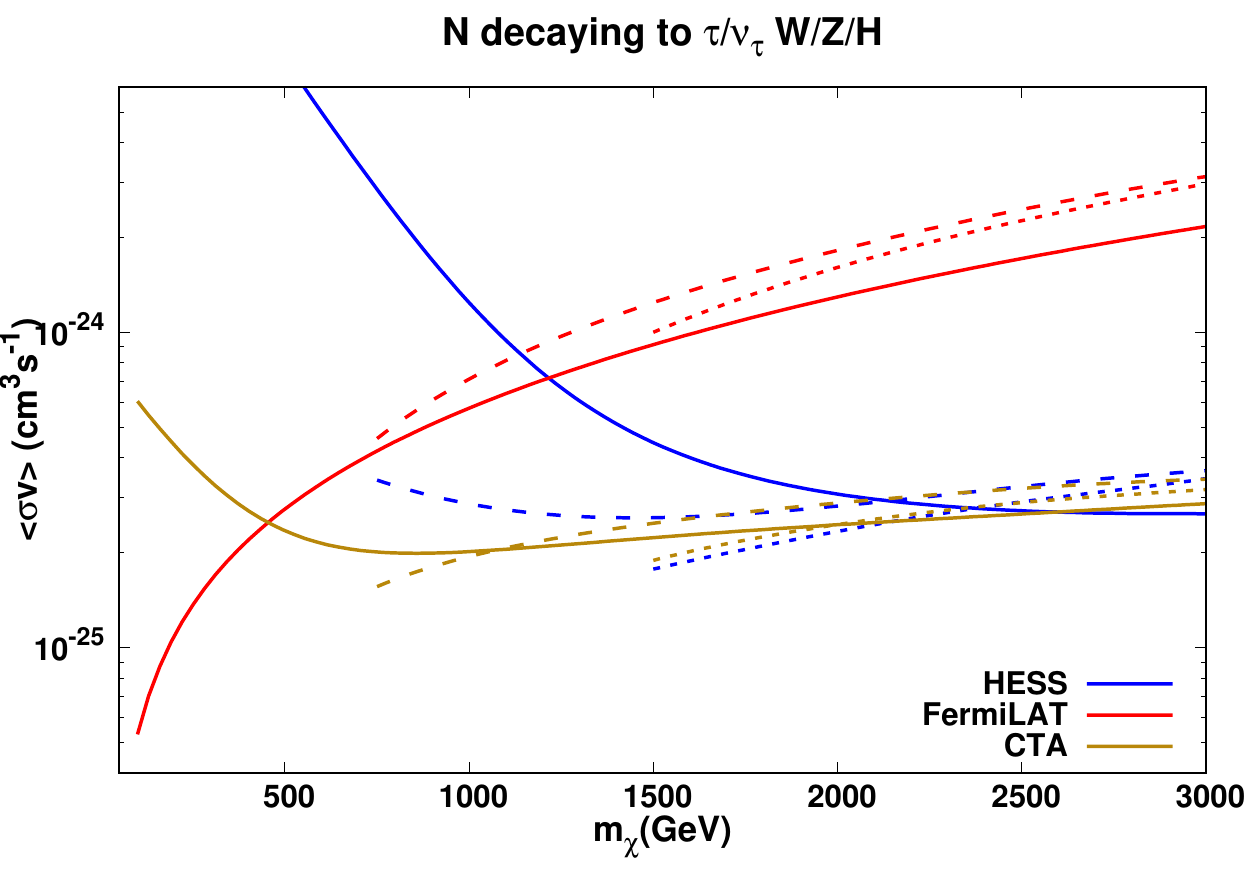}}}
\caption{\em Upper limits on $\langle \sigma v\rangle$ as a function
  of $m_\chi$ for a) electron-RHN final state and b) tau-RHN final
  state. We obtained these limits from different experiments
  Fermi-LAT, H.E.S.S. and CTA for different values of RHN mass,
  $m_N\,=\,100$ GeV (Solid), $m_N\,=\,750$ GeV (dashed) and
  $m_N\,=\,1500$ GeV (dotted). \label{fig:bounds}}
\end{figure}

The consequent constraints, in the $m_\chi$--$\langle\sigma v\rangle$
plane, from the Fermi-LAT and H.E.S.S. as also the projected CTA
sensitivity are depicted in \autoref{fig:bounds}. The area above the
curves are (would be) ruled out at 95\% C.L. Understandably, the
limits for the tau--channel are slightly more stringent as it leads to
a slightly larger gamma-ray production. As pointed out earlier, the
FermiLAT experiment is more sensitive at lower energies, whereas
H.E.S.S. is sensitive in the $0.2-30$ TeV range. Similarly, the energy
threshold of the CTA is lower than that of H.E.S.S., making the CTA
much more sensitive to photons of intermediate energies. Thus, CTA
would furnish the strongest bounds for dark matter masses above 600
GeV almost independent of $m_N$, whereas the FermiLAT gives better
limits for $m_\chi \lsim 500\gev$. For $m_\chi$
greater than a few TeV, H.E.S.S. already provides bounds equally
significant as as those the CTA would lead to.

The shape of the constraint curves are determined by a convolution of
the $\gamma$-ray spectrum with the energy sensitivity. To understand
the final outcome, several aspects have to be borne in mind:
\begin{itemize}
  \item As we have discussed earlier, the three detectors have peak
    sensitivity at different $E_\gamma$ ranges. Fermi-LAT is the most
    sensitive at lower energies (0.5 GeV to 0.5 TeV) while for
    H.E.S.S,, this lies in the few TeVs range. As for the CTA,
    although it covers a very wide range ($\sim$60 GeV to $\sim$300
    TeV), for low-energy (in tens of GeVs) $\gamma$-rays it obviously
    loses out to Fermi-LAT. For somewhat higher energies, it competes
    very will with the H.E.S.S., and goes much beyond.

  \item Thus, we would expect the Fermi-LAT to be the most sensitive detector
    for relatively low DM masses, as is borne out by \autoref{fig:bounds}.
    
  \item Naively, one would expect that a larger value for $m_N$ would,
    typically, translate to more energetic decay products and, hence, a
    harder $\gamma$-ray spectrum. Consequently, the constraints from
    non-observation would be stronger. This is reflected by the fact of the
    constraints being stronger for $m_N = 1.5$~TeV than they are for
    $m_N = 0.75$~TeV. This argument, though, fails for the $m_N = 100$~GeV case.

  \item For $m_N \ll m_\chi$, the RHN is highly boosted and this
    is transferred to its daughters, thereby allowing them to radiate
    off more photons. Some of these would have energies between 20 GeV
    and 0.5 TeV and visible to the detectors under discussion. With Fermi-LAT's
    sensitivity to low energy $\gamma$-rays being the highest, the improvement
    is most pronounced for this case.

    \end{itemize}

Compared to DM annihilation \cite{Campos:2017odj, Jangid:2020qgo}
these indirect bounds are less constrained for the semi-annihilation
processes \cite{semianniSca1}, as expected owing to less photon in the
final state.

We also analysed the sensitivity of semi-annihilating DM through their
neutrino spectrum in detectors like Super-kamiokande~\cite{Super-Kamiokande:2015qek}. We found that 
the limits are atleast $\mathcal{O}(1)$ magnitude weaker than the DM
annihilating to two neutrinos~\cite{Arguelles:2019ouk} and 2-3 orders of magnitude weaker than the complimentary photon channel discussed above. 
This is due to the  fact that  the  neutrino spectrum is weaker in comparison to photon spectrum and  additionally, neutrinos interact weakly with the detectors.  
Thus, a dedicated analyses that include direction and energy information might provide the stronger limits. Similarly, neutrino experiments/detectors
like DUNE~\cite{DUNE:2020ypp} and Hyper-Kamiokande~\cite{Olivares-DelCampo:2018pdl} can provide stronger limits for low-$\chi$ whereas IceCube~\cite{IceCube:2016oqp} and ANTARES~\cite{ANTARES:2015vis} can be utilized for heavy-$\chi$. Even so, these are expected to be weaker than the complementary photon channel.



Before end this section we would like to comment on the fact that 
semi-annihilation processes inn an effective theory does not have mediator to couple to the nucleons, thus  do  not put any direct detection constraints. However, for an UV complete theory as discussed in \autoref{appendix:A}
can lead to the restriction on nucleon-DM couplings.

\section{Conclusion}
\label{conc}
The continued absence of any signal for particulate Dark Matter, other
than the astrophysical or cosmological inferences calls for a
reevaluation of the paradigm. One particular aspect is the mechanism
that renders it stable against decay. While in most theories, this is
ensured by the imposition of a $Z_2$ symmetry, it certainly is not the
only avenue.  Alternatives such as $Z_n \, (n > 2)$, or even more
complicated ones, are not only feasible, but also dramatically alter
the expectations in both laboratory (direct detection) as well as
satellite-bound (indirect detection) experiments.

In this study we consider the simplest alternative, namely a $Z_3$
symmetry under which the DM $\chi$ transforms nontrivially while all the
Standard Model fields transform trivially. While similar efforts have
been made earlier for scalar DM, we eschew that path for every new
fundamental scalars brings along its own hierarchy problem.  Instead,
we choose to work with a fermionic DM.

We further augment the theory by the inclusion of right-handed (SM- as
well as $Z_3$ neutral) neutrino fields $N_i$. These serve twin purposes.
Allowed Majorana masses as well as nonvanishing Yukawa couplings
connecting with the SM neutrinos, they allow the generation of correct
masses for the light neutrinos (via the seesaw
mechanism). Furthermore, they offer an efficient avenue for the
disappearance of the DM so as to obtain the correct relic density.

Rather than adopt a particular UV-complete theory (although we do
offer one in an Appendix), we have chosen to work in the effective
theory framework (with a cutoff scale $\Lambda$)
through the inclusion of dimension-six operators
involving $\chi$ and the $N_i$. While complete annihilations
($\bar \chi + \chi \to N_i + N_j$) are possible, for a very wide range
of parameters (as exemplified in the Appendix), it is the
semi-annihilation process ($\chi + \chi \to \chi^c + N_i$) that
dominates. 

The Boltzmann equations governing the evolution of the DM density
changes from the usual paradigm not only because it is the
semi-annihilation that dominates, but also because it is coupled with
the evolution of the $N_i$ densities, which decay to the SM particles
through their Yukawa couplings.  If the size of these couplings are to
be commensurate with the observed neutrino masses, they do help in
keeping the $N_i$ in equilibrium with the SM plasma until relatively
late. The exact sizes of these couplings understandably have a very
big effect on the $N_i$ decoupling era and, hence, their densities at
that epoch.  However, the effect on the $\chi$ relic density is not as
pronounced. Given this, we have obtained the region in the EFT
parameter space that correctly reproduces the relic density for
various choices of the operators in the effective Lagrangian. On the
adoption of a specific UV-complete theory (as in
Appendix \ref{appendix:A}), this can be easily translated to a
corresponding statement for the said theory. 

While direct detection experiments would obviously be insensitive to
such a scenario, the indirect detection theatre is a very interesting
one.  The decays of the $N_i$ (themselves the products of the
semi-annihilation process) yield charged particles (gauge
bosons/quarks and charged leptons). These, as well as their cascade
decay products, radiate energetic photons which could be detected by
the satellite experiments like FermiLAT or earth based telescope like
the H.E.S.S. and CTA. We explored such detection possibilities in
detail, simulating the photon energy spectrum and convoluting with the
detector responses.  While non-observation of such signals the
Fermi-LAT already imposes interesting bounds, especially for a light
DM, future data from H.E.S.S.-II and the CTA are expected to lead to
strong bounds in the the heavy-$\chi$ region. As an example, for $m_\chi\sim\,100{\,\rm GeV}$, 
a thermal annihilation cross 
section  $\langle \sigma v\rangle\sim 5\times 10^{-26}{\rm cm}^3{\rm
s}^{-1}$ would be ruled out. For heavier $m_\chi$, the bounds are weaker,  
 and most of the parameter space satisfying relic abundance is
allowed by the Fermi-LAT, H.E.S.S and C.T.A data. And although, the neutrino spectrum of RHN's, can, in principle, be detected in 
several neutrino detectors,  the
corresponding constraints are much weaker
compared to those obtained from the gamma-ray telescopes.

With the effective theory presented here easily escaping
constraints from current and near-future experiments,
it is obvious that such a scenario can indeed alleviate the tension between the {\sc Planck} measurements of the DM relic density on the one hand and the null results from direct and indirect detections of DM on the other. It also needs to be appreciated that the quest for an UV complete theory  (such as
    that discussed in \autoref{appendix:A}) could engender DM-nucleon
    interactions   as well as self-interactions of the DM. While the former would open new windows to both direct and indirect
    detection, the latter could be expected to leave a discernible mark on the
    small-scale structure in galaxies and clusters. The paradigm, thus, holds
    much promise.

\section*{Acknowledgements}
PB wants to thank Anomalies 2019, 2020, SERB CORE Grant
CRG/2018/004971 and MATRICS Grant MTR/2020/000668 for the financial
support. PB also acknowledges Farinaldo Queiroz for the valuable
comments. DS has received funding from the European Union's Horizon 2020
research and innovation programme under grant agreement No 101002846, ERC
CoG CosmoChart.
\begin{appendices}
\section{An ultraviolet completion}
\label{appendix:A}
While the Lagrangian of (\autoref{dim-6}) is quite acceptable as an
effective theory, a renormalizable theory is always
desirable. Furthermore, the absence of certain terms (such as
$\bar\chi \Gamma_i \chi \, \bar N \Gamma_j N$, with $\Gamma_{i,j}$
being commensurate Dirac matrices) in (\autoref{dim-6}) is a cause of
concern for their presence would tend to reduce the efficacy of
semi-annihilation. In this Appendix, we present a possible ultraviolet
completion that addresses both these issues.

To this end, let us introduce, to the model in \autoref{model}, a
$SU(2)_L$ singlet neutral scalar $\phi$ which transforms, under $Z_3$,
just as $\chi$ does. This would allow for Yukawa terms of the form
\begin{equation}
  L_{int} \supset \bar{\chi^c} \left(y_{2} + y_{2}' \gamma_5\right) \chi \phi
  +   \phi\, \bar \chi \left(y_{3} + y_{3}' \gamma_5\right) \, N
  + \text{H.c.} 
\label{dmc}
\end{equation}
In the event of a heavy $\phi$, the field can be integrated out to yield
effective four-fermion vertices of the form $(C_{ab} / m_\phi^{2}) \, \left(\bar N \Gamma_a \chi \right)
                    \left(\bar{\chi^c} \Gamma_{b} \chi\right)$
with the coefficients $c_{ab}$ being related through Fierz rotations.

Of course, analogous terms such as $\bar\chi \Gamma_i \chi \, \bar N \Gamma_j N$ would be generated as well. However, for $|y_3, y_3'| \ll |y_2, y_2'|$ ---
a technically natural choice --- such operators are relatively
suppressed with the consequence that semi-annihilation wins over annihilation.

It is instructive to consider the potential for the scalars in
the theory, namely $\phi$ and the usual SM doublet $\Phi$. The most
general form is given by
\begin{equation}
\begin{aligned}
V(\Phi, \phi)=  \mu_1^2|\Phi|^2 +\mu_2^2 |\phi|^2+\frac{\mu_3}{3} \left(\phi^3 + \phi^{\dagger 3}\right) +\frac{\lambda_1}{2}|\Phi|^4+ \frac{\lambda_2}{2} |\phi|^4 + \lambda_{\phi}|\phi|^2 |\Phi|^2 +
 \text{H.c.}    
\end{aligned}
\label{vphi}
\end{equation} 
A nonzero value for $\langle\phi\rangle$ would result in $\phi-\Phi$
mixing, and thereby to direct $\chi \chi$ annihilation to the SM
particles through the Higgs portal. However, the spontaneous breaking
of $Z_3$ that this entails would also allow the DM to decay. While the
lifetime could be suitably extended by choosing parameters
appropriately, the solutions tend to be somewhat unnatural and we
eschew that path.

\section{Getting to the cross sections}\label{appendix:B}
We list below the squares of the matrix
elements for the process in \autoref{the_process} using the
Lagrangian of  \autoref{dim-6} with the assumption that only one of
the $c_{ab}$ is non-zero and set to unity. Defining the Lorentz-invariant
quantities
\begin{equation}
\begin{array}{rcl}
{\cal A}_4 & \equiv & \dis q_1 \cdot q_3 \, q_2 \cdot q_4 + q_1 \cdot q_2 \, q_3 \cdot q_4 + q_1 \cdot q_4 \, q_2 \cdot q_3
\\[1ex]
{\cal A}_2 & \equiv & \dis q_1 \cdot q_3 + q_2 \cdot q_3 + q_3 \cdot q_4
\\[1ex]
\widetilde {\cal A}_2 & \equiv & \dis q_1 \cdot q_2+q_1 \cdot q_4+q_2 \cdot q_4
\end{array}
\end{equation}
we have
\begin{equation}
\begin{array}{rcl}
\dis |M|^2_{SS} = |M|^2_{PP} 
 & = & \dis\frac{8}{\Lambda^4}\left( {\cal A}_4 -m^2\,{\cal A}_2
 + mm_N\,\widetilde{\cal A}_2 -3m^3m_N\right)
 \\[2ex]
|M|^2_{SP}\,= |M|^2_{PS}\,&=&\,\dis\frac{8}{\Lambda^4} \left(-{\cal A}_4
+m^2\,{\cal A}_2 + mm_N\,\widetilde{\cal A}_2-3m^3m_N\right) \\[2ex]
|M|^2_{VV}\,&=&\,\dis\frac{128}{\Lambda^4}  \left({\cal A}_4 +\frac{ m^2}{2}\,{\cal A}_2  - \frac{mm_N}{2}\,\widetilde{\cal A}_2-3m^3m_N\right) \\[2ex]
|M|^2_{VA}\,&=&\,\dis\frac{128}{\Lambda^4}  \left({\cal A}_2 -m^2\,{\cal A}_2
- mm_N\,\widetilde{\cal A}_2+3m^3m_N\right) \\[2ex]
  |M|^2_{AA}\,&=&\,\dis\frac{128}{\Lambda^4}  \left({\cal A}_4 -m^2\,{\cal A}_2
   + mm_N\,\widetilde{\cal A}_2- 3m^3m_N\right) 
\end{array}
\end{equation}
Note that spin-summing (and averaging) is yet to be done. The
corresponding expressions for the reverse process
($\bar \chi^c \to \chi \chi$) or analogous ones (such as $\chi
N \to \chi^c \chi^c$ {\em etc.}) can be obtained from those above
using crossing symmetry.

Similarly, the decay widths for the RHN are given by
\begin{equation}\label{pardcy1}
\begin{array}{rclcl}
\dis	\Gamma(N_i \to  W^{+}l^{-})
& = & \dis \Gamma(N_i \to  W^{-}l^{+})&=& \dis
\frac{y_N^2 M_{N_i}}{32 \pi} \left( 1- \frac{M_W^2}{M_{N_i}^2}\right)^2 \left(1+\frac{2 M_W^2}{M_{N_i}^2}\right),\\[15pt]
\dis	\Gamma(N_i \to Z\, \nu)& = & \dis
\Gamma(N_i \to Z\, \bar{\nu})&=& \dis
\frac{y_{N_i}^2 M_{N_i}}{64 \pi} \left( 1- \frac{M_Z^2}{M_{N_i}^2}\right)^2 \left(1+\frac{2 M_Z^2}{M_{N_i}^2}\right).
	\\[15pt] 
\dis	\Gamma(N_i \to h\, \nu)& = & \dis \Gamma(N_i \to h\, \bar{\nu})&=& \dis
\frac{y_{N_i}^2 M_{N_i}}{64 \pi} \left( 1- \frac{M_h^2}{M_{N_i}^2}\right)^2. 
\end{array}
\end{equation}
where the masses of the charged leptons have been neglected.

\section{Boltzmann Equation}\label{appendix:C}
We, now, derive the Boltzmann equation relevant for the Dark Matter
density evolution in the present context. With $2 \to 2$ scattering
being the dominant process, the calculation proceeds quite similarly
to the standard case, except for the fact that only one DM particle is
produced (or destroyed) per collision process. On the other hand,
there would be two identical particles in either the initial or the
final state, thereby necessitates the inclusion of a factor of
$(1/2!)$ in the phase space calculation. This gives, for the $1 +
2 \to 3 + 4$ process, the time-variation of the number density
$n_\chi$ to be
\begin{equation}
\begin{array}{rcl}
\dis \frac{dn_\chi}{dt}\,+\,3Hn_\chi\,& = & \dis \,\frac{-1}{2!} 
 \int \left[d {\cal P}\right] 
\Big(|M_{12\to 34}|^2f_1f_2-|M_{34\to 12}|^2f_3f_4\Big)
\end{array}
\label{eq:boltz}
\end{equation}
where, for the sake of simplicity, we have suppressed the Pauli
blocking\footnote{For non-relativistic and heavy particles, the
blocking is, anyway, numerically unimportant.} factors.  Here, $f_i$s
represent the appropriate statistical distribution factors, $H$ is the
instantaneous Hubble expansion rate and
\begin{equation}
[d {\cal P}] \equiv \left(\prod_{i = 1}^4 \frac{d^3p_i}{(2\pi)^32E_i} \right) \,
                (2\pi)^4\delta^4(p_1+p_2-p_3-p_4)
\end{equation}
with appropriate spin-sum and averaging being understood.
If we assume time-reversal invariance, we have $|M_{12\to 34}|^2 = |M_{34\to
12}|^2$ and, thus, the right hand side of eqn. \ref{eq:boltz} can be written as
\be
\text{R.H.S}\,=\,\frac{-1}{2}\int  \left[d {\cal P}\right] 
   |M_{12\to 34}|^2 \left(f_1f_2-f_3f_4\right) \ .
\ee
Rather than continue with the number density $n_\chi$, it is customary to
consider the {\em yield} which is defined as it ratio with
the ambient entropy density $s$, {\em viz.} $Y_\chi \equiv n_\chi / s. $
This immediately leads to 
\[\frac{dn_\chi}{dt}\,+\,3Hn_\chi\,=\,s\frac{dY_\chi}{dt} \ .
\]
Noting that, during the cosmological evolution, there exists
a monotonic relation between the time elapsed and the temperature of the
universe, it is useful to consider a  change of variables
\[x=\frac{m_\chi}{T}\] 
such that we have
\be
\frac{dY_\chi}{dx}=\frac{-x}{2H(m_\chi)s}\int \left[d {\cal P}\right] 
\ee
where
\[
H(m_\chi)\,=\,\displaystyle\frac{\pi\sqrt{g_*}m_\chi^2}{\sqrt{90}M_\text{pl}}
\qquad {\rm and} \qquad 
s\,=\,\displaystyle\frac{2\pi^2g_*T^3}{45}\,=\,\frac{2\pi^2g_*m_\chi^3}{45x^3}
\ .
\]
with $g_*$ denoting the number of degrees of freedom relevant at that
temperature.

It is an excellent approximation to consider these particles $(N$ and $\chi$)
to be in kinetic equilibrium during the freeze out. This allows us to write
\[f(E,t)=\frac{n(x)}{n^\text{eq.}(x)}f^\text{eq.}(E,t)\]
where
\[{f^\text{eq.}\,=\,\frac{1}{e^{E/T}+1}\qquad\text{and}\qquad n^\text{eq.}\,=\,\displaystyle\frac{g}{2\pi^2}\int_m^\infty\frac{(E^2-m^2)^{1/2}}{e^{E/T} +1 }E\,dE}
\]
both $\chi$ and $N_i$ being fermionic. Now, decoupling occurs, typically, at $x \equiv m_\chi/T \gsim 20$
and, hence, in $f^\text{eq}(\chi)$, we may safely approximate
$(e^{E/T} + 1)^{-1} \approx e^{-E/T}$. And, since we would not be
contemplating a large hierarchy between $m_\chi$ and $m_N$, an analogous
approximation would hold as well for $f^\text{eq}(N_i)$.
Furthermore, using conservation of energy, we have
\begin{equation}
f^\text{eq.}_1(x)f^\text{eq.}_2(x) \approx e^{-(E_1+E_2)/T}\,=\,e^{-(E_3+E_4)/T}
\approx f^\text{eq.}_3(x)f^\text{eq.}_4(x)
\end{equation}
This leads to
\begin{equation}
\begin{array}{rcl}
\dis \frac{dY_\chi}{dx}&=&\dis \frac{-x}{2H(m)s}
\int \left[d {\cal P}\right] |M_{12\to 34}|^2 
\Bigg[\frac{n_1(x)n_2(x)}{n^\text{eq.}_1(x)n^\text{eq.}_2(x)}f^\text{eq.}_1(E,t)f^\text{eq.}_2(E,t) \\[3ex]
&&\dis \hspace*{12em} -
\frac{n_3(x)n_4(x)}{n^\text{eq.}_3(x)n^\text{eq.}_4(x)}f^\text{eq.}_3(E,t)f^\text{eq.}_4(E,t)\Bigg]
\\  [4ex]
&=&\dis\frac{-x}{2H(m)s}\int \left[d {\cal P}\right] \frac{|M_{12\to 34}|^2}
{n^\text{eq.}_1(x)n^\text{eq.}_2(x)}
   f^\text{eq.}_1(E,t)f^\text{eq.}_2(E,t)\nonumber \\[3ex]
&&\dis \hspace*{6em} \left[n_1(x)n_2(x)-
       n_3(x)n_4(x)\frac{n^\text{eq.}_1(x)n^\text{eq.}_2(x)}
                        {n^\text{eq.}_3(x)n^\text{eq.}_4(x)}\right]
\\[4ex]
&=&\displaystyle\frac{-x\,s}{2H(m)}\int \left[d {\cal P}\right]
\frac{|M_{12\to 34}|^2}
{n^\text{eq.}_1(x)n^\text{eq.}_2(x)}f^\text{eq.}_1(E,t)f^\text{eq.}_2(E,t)\nonumber
\\[3ex]
&&\dis \hspace*{6em} \left[Y_1(x)Y_2(x)-
Y_3(x)Y_4(x)\frac{Y^\text{eq.}_1(x)Y^\text{eq.}_2(x)}
                 {Y^\text{eq.}_3(x)Y^\text{eq.}_4(x)}\right] \ .
\end{array}
\end{equation}
Thus, defining a thermal average as
\[
\langle\sigma v\rangle \equiv
\int \left(\prod_{i=1}^4 d\Pi_i\right) \frac{(2\pi)^4\delta^4(p_1+p_2-p_3-p_4)}{n^\text{eq.}_1(x)n^\text{eq.}_2(x)}|M_{12\to 34}|^2f^\text{eq.}_1(E,t)f^\text{eq.}_2(E,t)
\ ,
\]
we obtain the set of equations describing the evolution of 
number density of DM as well as RHN in the early universe:
\begin{equation}
\begin{array}{rcl}
 \dis   \frac{dY_{\chi}}{dx}& =& \dis \frac{- x\,s(m_\chi)}{2H(m_\chi)}
    \langle\sigma v\rangle^{\chi \chi \rightarrow \chi N}
     \left[Y_{\chi}^2- \frac{Y^{Eq}_{\chi}}{Y^{Eq}_{N}} Y_{\chi}\,Y_{N}\right]
     \\[4ex]
\dis
\frac{dY_{N}}{dx}& =& \dis
\frac{x\,s(m_\chi)}{2H(m_\chi)} \langle\sigma v\rangle^{\chi \chi \rightarrow \chi N}
\left[ Y_{\chi}^2-\frac{Y^{Eq}_{\chi}}{Y^{Eq}_{N}} Y_{\chi}\,Y_{N}\right]
    -\frac{\Gamma_{N}\,x}{H(m_\chi)} (Y_{N}-Y_{eq}) \ .
       \label{eq_Boltz}
\end{array}
\end{equation}
Note that the second term in $dY_{N}/dx$ carries the information of the
RHN decaying into the SM particles.

\end{appendices}

\bibliographystyle{unsrt}

\end{document}